\documentclass[
  journal=pasa,
  manuscript=article-type,
  year=2020,
  volume=37,
]{cup-journal}
\usepackage{hyperref}
\hypersetup{
    colorlinks = true,
    allcolors  = {blue} 
} 
\usepackage{amsmath}
\usepackage{amssymb}
\usepackage[nopatch]{microtype}
\usepackage{booktabs}
\usepackage{natbib}
\definecolor{darkgreen}{rgb}{0.0, 0.5, 0.0}
\newcommand{\msun}[1]{M$_\odot$}

\title{Second Generation Planet Formation in Post-AGB discs -- II: Dust Coagulation and Core Accretion}

\defcitealias{pourmand2025secondgenerationplanetformationpostagb}{Paper~I}

\author{Ali Pourmand}
\affiliation{School of Mathematical and Physical Sciences, Macquarie University, Sydney, NSW, Australia}
\alsoaffiliation{Astrophysics and Space Technologies Research Centre, Macquarie
University, Sydney, NSW, Australia}
\author{Orsola De Marco}
\affiliation{School of Mathematical and Physical Sciences, Macquarie University, Sydney, NSW, Australia}
\alsoaffiliation{Astrophysics and Space Technologies Research Centre, Macquarie
University, Sydney, NSW, Australia}
\author{Devika Kamath} 
\affiliation{School of Mathematical and Physical Sciences, Macquarie University, Sydney, NSW, Australia}
\alsoaffiliation{Astrophysics and Space Technologies Research Centre, Macquarie
University, Sydney, NSW, Australia}
\author{Daniel J. Price}
\affiliation{School of Physics and Astronomy, Monash
University, Melbourne, VIC, Australia}

\email[Ali Pourmand]{ali.pourmand@hdr.mq.edu.au}

\keywords{stars: AGB and post-AGB, planets and satellites: formation, protoplanetary discs} 

\begin{document}

\begin{abstract}
Observations have established that the circumbinary discs formed around post-asymptotic giant branch (post-AGB) binaries have striking similarities to protoplanetary discs around young stars. In this study, we examine the feasibility of the different stages of growth in the core accretion framework of planet formation in post-AGB discs, in order to assess whether planet formation is possible in these systems. We find that dust coagulation up to mm sizes is possible within the lifetimes of post-AGB discs, consistent with observations. We then investigate the subsequent growth of planetesimals through pebble and planetesimal accretion. If the streaming instability operates in these discs, the resulting planetesimals formed in higher-mass post-AGB discs ($M_{\rm disc}\sim0.1M_\odot$) can enter rapid pebble accretion and reach their isolation mass within the estimated disc lifetime. By contrast, in lower-mass post-AGB discs ($M_{\rm disc}\sim0.01M_\odot$), planetesimal accretion is too inefficient to produce substantial further growth. We also show that the larger aspect ratios of post-AGB discs compared to protoplanetary discs around young stars imply high pebble isolation masses and therefore potential formation of rocky planets with masses extending up to Jupiter masses. After assessing gas accretion and concurrent migration in these discs, we conclude that second generation planet formation in post-AGB discs is theoretically possible within their estimated lifetimes ($10^4$--$10^5$\,yr) provided that local dust-to-gas ratio enhancements and sufficiently high disc masses are achieved.
\end{abstract}

\section{Introduction}

\label{sec:intro}

More than 80 circumbinary discs around post-asymptotic giant branch (post-AGB) binaries have been catalogued in the Milky Way \citep[e.g.][]{kamath_2019,Kluska_2022,Andrych_2023}, with additional systems identified in the Small and Large Magellanic Clouds \citep{2014Kamath,Kamath_2015}. These systems typically contain a post-AGB primary, a main-sequence companion, and a circumbinary disc, inferred initially from the characteristic near-IR excess in their spectral energy distributions \citep{2006deRuyter,2014Kamath,Kamath_2015,Kluska_2022}. The discs are thought to form as a consequence of late-stage binary interaction and mass loss, although the details of their formation remain poorly understood \citep[e.g.][]{2017Chen,2021Malfait,2026Huang}. Based on their near- and mid-IR colours, \citet{Kluska_2022} classified post-AGB discs into several categories, including "full discs", in which dust extends inward to approximately the sublimation radius, and "transition discs", in which the inner dust rim lies significantly farther out (e.g. \citealt{1989Strom,1990Skrutskie,2014Espaillat}).

A growing body of observations has revealed striking similarities between post-AGB circumbinary discs and protoplanetary disc (PPDs) around young stellar objects (YSOs). Post-AGB discs show stable or quasi-Keplerian rotation \citep{Bujarrabal_2015,Gallardo_Cava_2023,2006deRuyter,Gallardo_Cava_2021}, evidence for grain growth and dust evolution \citep{2005deruyter,Scicluna_2020,Kluska_2022,Andrych_2023}, and differences between the radial extent of the gaseous and dusty components that may be indicative of radial drift \citep{bujarrabal2023compactdustdiskred}. Many systems also show strong photospheric depletion of refractory elements \citep{Kluska_2018,kamath_2019,mohorian2024measurementscarbonisotopicratios,mohorian2025tracingchemicaldepletionevolved}. One proposed explanation for this depletion is the preferential trapping of refractory-rich dust within the disc, potentially by pressure structures associated with a planet.


Post-AGB discs also differ from PPDs around YSOs in several important respects. Their estimated lifetimes are much shorter, of order $\sim10^5$ yr compared with $>10^6$ yr for typical PPDs \citep{Bujarrabal_2017,martin2025modellingimpactcircumbinarydisk}. They also have comparatively higher temperatures, implying larger sublimation radii \citep{Kluska_2019}.
Additional differences include the influence of binary--disc interactions and distinct initial gas and dust conditions. Nevertheless, the structural and physical similarities raise the possibility that post-AGB circumbinary discs may provide environments for ``second-generation'' planet formation.

Theoretical models of planet formation have primarily been developed in the context of PPDs around YSOs. Two principal frameworks are generally considered: gravitational instability and core accretion \citep[see review by][]{drazkowska2023planetformationtheoryera}. Gravitational instability is a top-down process in which sufficiently over-dense regions of a disc collapse under their own gravity to form bound clumps that may subsequently contract into planetary-mass objects \citep[e.g.][]{1997boss,2002mayer}. Core accretion \citep{1972Safronov,1996Pollack}, by contrast, is a bottom-up process in which small solid grains grow through collisions and other mechanisms to form planetesimals, which can then undergo further growth through gravitational accretion.

Although these frameworks were developed mainly for planetary systems around young stars, planets or planetary candidates have also been identified in evolved stellar environments. Examples include the pulsar planets \citep{1992Wolszczan,Bailes_2011}, the planetary-mass companion candidate in the L2~Puppis system, an AGB star surrounded by an edge-on circumbinary disc \citep{Kervella_2016}, and the two proposed circumbinary planets around NN~Serpentis, a post-common-envelope binary containing a white dwarf and a main-sequence companion \citep{2010Beuermann}. The origin of such planets remains uncertain, and both first- and later-generation formation scenarios have been discussed \citep[e.g.][]{1993Podsiadlowski,2014schleicher,pourmand2025secondgenerationplanetformationpostagb}.

In a preceding study (\citealp{pourmand2025secondgenerationplanetformationpostagb}; hereafter \citetalias{pourmand2025secondgenerationplanetformationpostagb}), we showed that gravitational instability is unlikely to form planets in post-AGB discs because of the combination of their relatively low disc masses and high temperatures ($T=1200$~K at $r\approx2$--$5$~au). In this work, we instead examine whether the successive stages of the core-accretion pathway can operate under the physical conditions and within the short lifetimes of post-AGB discs. In particular, we assess dust growth, planetesimal formation, pebble and planetesimal accretion, gas accretion, and migration to determine whether second-generation planet formation is theoretically feasible in these systems.

This paper is structured as follows: In Section~\ref{sec:observations} we review the observational evidence relevant to planet formation in post-AGB discs. In Section~\ref{section:coreaccretion}, we consider the successive stages of core accretion and evaluate whether each can operate under post-AGB disc conditions. We discuss the implications of our results and summarise our conclusions in Section~\ref{sec:summary}. 

\section{Observational Signatures of Post-AGB Discs Relevant to Planet Formation}
\label{sec:observations}
\begin{table*}[t]
\resizebox{\textwidth}{!}{%
\begin{threeparttable}
\caption{Post-AGB discs with observational features relevant to the presence of planets or ongoing stages of planet formation.}

\label{tab:post-agb-list}
\begin{tabular}{lllllll}
\toprule
 \headrow System / Signature 
& Transition disc
& Substructure 
& Inner Cavities
& $R_{\rm gas}/R_{\rm dust}$ 
& $M_{\rm gas}/M_{\rm dust}$ 
& Misalignment of \\
&  
&  
& 
&  
& 
& disc and Binary Orbit \\

\midrule
AC Her& yes 
& yes 
& 25 au
& - 
& $10^{a}$ 
& $9^{\circ}{}^{b}$ \\

CT Ori 
& yes 
& - 
& 5 au 
& - 
& - 
& - \\

EP Lyr 
& yes 
& - 
& 7 au 
& - 
& - 
& - \\

AD Aql 
& yes 
& - 
& 18 au 
& - 
& - 
& - \\

RU Cen 
& yes 
& - 
& 4 au 
& - 
& - 
& - \\
ST Pup 
& yes 
& - 
& 3 au 
& - 
& - 
& - \\

AR Pup 
& no 
& yes 
& - 
& - 
& - 
& - \\

U Mon 
& no 
& yes 
& - 
& - 
& - 
& - \\

IRAS 08544-4431 
& no 
& yes 
& - 
& - 
& - 
& maybe$^*$\\

W Car 
& - 
& yes 
& - 
& - 
& - 
& - \\

IRAS 15469-5311 
& - 
& yes 
& - 
& - 
& - 
& - \\

Red Rectangle 
& - 
& - 
& - 
& $\sim10^{c}$ 
& - 
& - \\

\bottomrule
\end{tabular}
\begin{tablenotes}[hang]
\footnotesize
\item[] Notes: The systems we have identified as "transition disc" are those classified as such by \citet{Kluska_2022}. The column "inner cavities" only includes inner cavity sizes of systems reported in \citep{Corporaal2023} whose inner cavity sizes were found to be larger than their respective sublimation radii. The column "substructure" are systems found to have substructure by \citet{Andrych_2023}; specifically they find that IRAS 08544-4431 has a ring-like substructure and IRAS 15469-5311 has a ring-shaped substructure. The superscripts denote the sources the parameters were taken from which are: $^a$ \citet{Hillen_2015}, $^b$ \citet{Anugu_2023}, $^c$ \citet{bujarrabal2023compactdustdiskred}. 

\item[] *: Further observations are required to determine whether the data indicate a misalignment between the disc and the binary orbit, or instead a warp within the disc \citep{Andrych_2024}.
\end{tablenotes}
\end{threeparttable}%
}
\end{table*}

Specific observational signatures possibly relevant to planet formation are described in \ref{ap:obs_features_pagb} (for a list of relevant individual systems see Table~\ref{tab:post-agb-list}). Here, we instead highlight the key disc properties relevant to planet formation and their characteristic values in post-AGB discs and PPDs around YSOs.

The first disc parameter critical to planet formation is the aspect ratio, $H/r$, which is larger in post-AGB discs than in typical PPDs around YSOs. It can be expressed as

\begin{equation}
    \label{eq:aspectratio}
    \frac{H}{r}\approx0.12\left(\frac{L_1}{5.61\times10^{3}L_\odot}\right)^{3/16}\left(\frac{M_{\rm cnt}}{1.5M_{\odot}}\right)^{-1/2}\left(\frac{r}{\rm 4.03au}\right)^{1/8},
\end{equation}
where $H$ is the scale height, $L_1$ is the luminosity of the primary star (the value $5.61\times10^{3}$~L$_{\odot}$ is that of a post-AGB star with $R_{\rm star}=100$~R$_\odot$, $T_{\rm star}=5000$~K), and $M_{\rm cnt}=M_1+M_2$ is the total mass of the two stars in the system. 
            
This expression is derived based on the following considerations: radiative-transfer modelling of post-AGB discs indicates that their midplane temperatures follow a steep radial decline $T \propto r^{q}$ consistent with power-law indices $q \sim -0.9$ to $-0.5$. For example, modelling of IRAS 08544-4431 yields midplane slopes of $-0.89<q<-0.75$ \citep{Corporaal_2023}, with the inner rim coinciding with the dust sublimation radius, inside which temperatures exceed the dust sublimation temperature, $T_{\rm sub}=1200$~K \citep{Kluska_2018,Kluska_2019}.

Additionally, interferometric size comparisons of several post-AGB discs from MIDI and PIONIER show that $-0.5>q>-0.75$ \citep{Kluska_2019}. We therefore adopt $q=-0.75$, consistent with the temperature profile inferred by \citet{Corporaal_2023}, although observed temperature profiles in YSO discs can be considerably shallower \citep[e.g.,][]{Dullemond2020}. This value is also consistent with the theoretical expectation for an optically thick, flat disc in radiative equilibrium \citep[e.g.,][]{Chiang_1997}. The resulting temperature profile is given by
\begin{equation}
\label{eq:Tprofile}
\begin{aligned}
T(r) = 1200K \left(\frac{r}{R_{\rm sub}}\right)^{-0.75},
\end{aligned}
\end{equation}
where the sublimation radius is
\begin{equation}
\label{eq:sublim}
    R_{\rm sub} \approx \frac{1}{2}\sqrt{\frac{L_{1}}{4\pi\sigma T_{\rm sub}^4}}=4.03{\rm au}\left(\frac{L_{1}}{5.61\times10^{3}L_{\odot}}\right)^{0.5},
\end{equation}
with $T(r=R_{\rm sub})=T_{\rm sub}=1200$~K. The sublimation radius in these discs is much larger than for YSOs, due to the high luminosity of post-AGB stars. 

Using the temperature profile we can estimate the sound speed

\begin{equation}
\label{eq:sound_speed}
c_{\rm s}(r)=\sqrt{\gamma \frac{k_{\rm B}T(r)}{\mu m_{\rm p}}},
\end{equation}
where $k_{\rm B}$ and $m_{\rm p}$ are the Boltzmann constant and proton mass, respectively, and the adiabatic index is set to $\gamma=1$ for an isothermal gas. We also adopt a mean molecular weight of $\mu=2.3$ appropriate for a predominantly molecular gas with a solar composition. 

\begin{figure}[t]
\centering
\includegraphics[width=\linewidth]{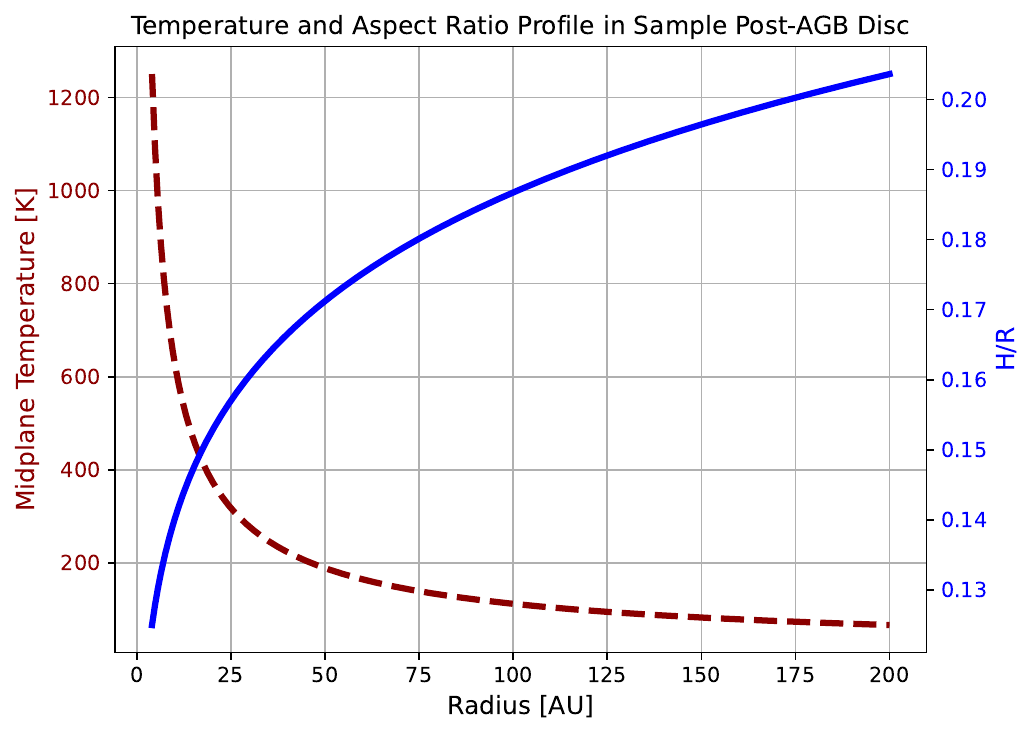}
\caption{Temperature and aspect ratio profiles predicted by Equations~\eqref{eq:Tprofile} and \eqref{eq:aspectratio}, respectively, for a representative post-AGB disc with $R_{\rm star}=100R_\odot$, $T_{\rm star}=5000$~K, and $M_{\rm cnt}=1.5M_\odot$.}
\label{fig:aspect_temp}
\end{figure}

With the sound speed in hand, the scale height and hence the aspect ratio \footnote{We note that using the temperature profile reported in Figure~10 of \citet{Corporaal_2023} we cannot reproduce the scale height fit they find; this is due to the different parametrisation used by their code to define the disc's scale height.} can be calculated using $H(r)=\frac{c_s(r)}{\Omega_p(r)}$, where $\Omega_P=\Omega_K=\sqrt{GM_{\rm cnt}/r^3}$ is the Keplerian angular frequency. The resulting aspect-ratio and temperature profiles for our representative post-AGB disc model are shown in Figure~\ref{fig:aspect_temp}. 

Following \citetalias{pourmand2025secondgenerationplanetformationpostagb} we take the surface density of the disc to be a power law $\Sigma(r) \propto r^{-n}$  with an inner radius equal to $R_{\rm sub}$, and an outer truncation radius $R_{\rm out}$. The gas and dust surface densities are assumed to follow a fixed dust-to-gas ratio. The normalisation for the surface density is found by setting an outer radius and a disc mass $M_{\rm disc}$. Post-AGB disc mass estimations vary depending on the measurement method: molecular line observations (e.g., CO) with ALMA suggest masses of $10^{-4}$--$10^{-2}$~M$_\odot$ \citep[e.g.,][]{Bujarrabal_2013,Gallardo_Cava_2021}, whereas radiative transfer modelling of dust emission from data from observations in the near and mid-infrared can yield higher values of 0.1--0.2~M$_\odot$ in several cases \citep{Kluska_2018,Corporaal_2023}. These estimates are sensitive
to assumptions about the dust's optical depth, grain properties, the dust-to-gas ratio, the vertical structure, and the disc's inclination, and are subject to model degeneracies.

Finally, the estimated lifetime of a post-AGB disc is $\lesssim 10^5$~yrs \citep{Bujarrabal_2017,martin2025modellingimpactcircumbinarydisk}, approximately an order of magnitude shorter than the typical lifetime of protoplanetary discs around YSOs.
With these characteristic values and relations established, we now assess the successive stages of core accretion to determine whether planet formation can proceed within post-AGB discs.

\section{The Stages of Planet Formation by Core Accretion}
\label{section:coreaccretion}

\begin{figure*}[t]
\centering
\includegraphics[width=\textwidth]{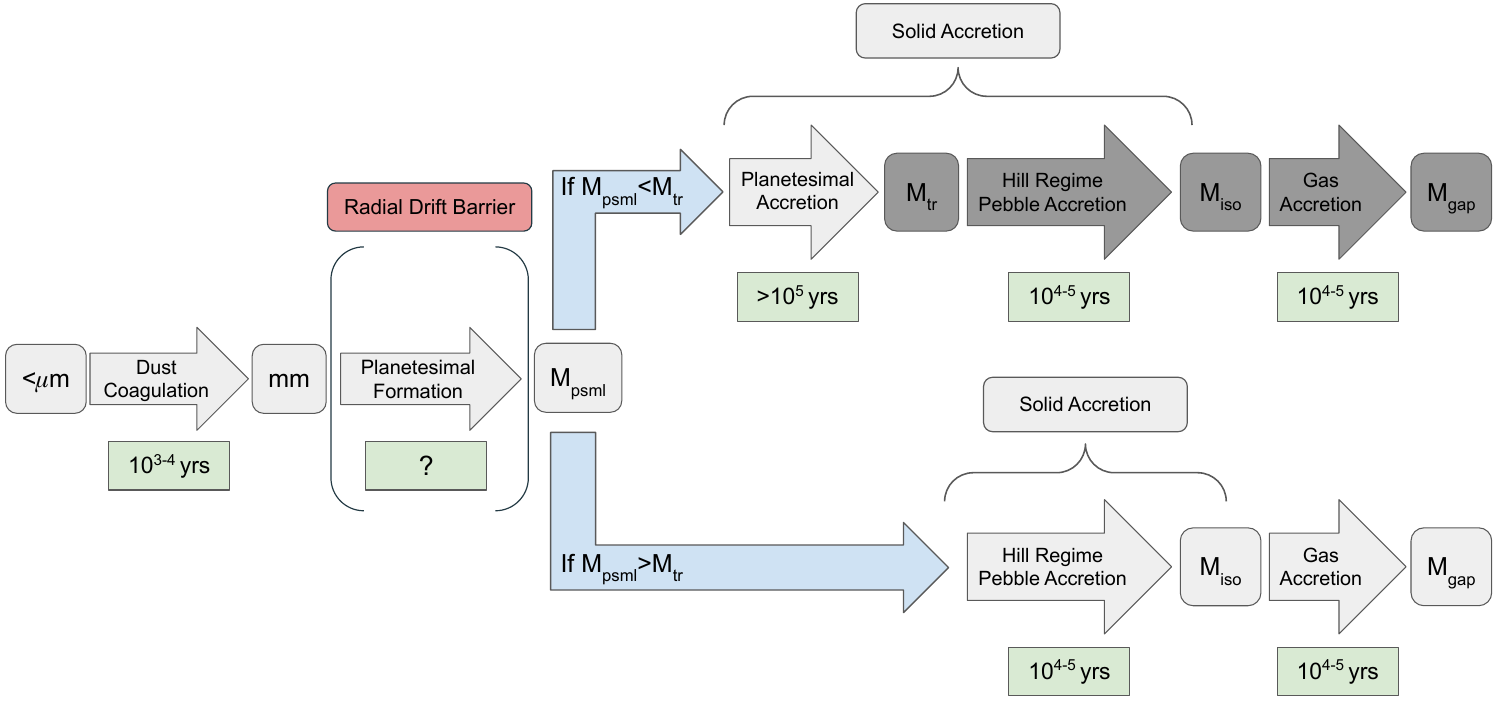}
\caption{Schematic depicting possible growth path up to planet sizes in post-AGB discs. Timescales written down are characteristic, and assuming the post-AGB disc has a mass of $0.01M_\odot$ (growth timescale written for Hill regime is also assuming local dust-to-gas ratio of $0.1-1$). The dark shaded boxes highlight unlikely processes because of excessive timescales. }
\label{fig:growthpath}
\end{figure*}

The bottom-up core accretion framework of planet formation consists of a sequence of growth stages \citep[seem e.g.,][]{drazkowska2023planetformationtheoryera}, illustrated in Figure~\ref{fig:growthpath} for reference. In the {\it Dust Coagulation} stage, sub-micron grains inherited from the ISM grow through collisions, leading to grains sticking to each other, bouncing off each other or fragmenting depending on conditions \citep{2001Weingartner,OriginsDodson2021,2000blum_experiments}. As particles grow, several growth barriers limit further coagulation \citep{Blum_2018,2010guttler_zoo}. Subsequent growth and {\it Planetesimal Formation} ($\sim 100-1000$~km) can proceed through the gravitational collapse of locally over-dense solids, potentially via mechanisms such as the {\it Streaming Instability} \citep{Youdin_2005}.

Once formed, planetesimals are thought to initially grow through mutual collisions ({\it Planetesimal Accretion}), a relatively slow process that is thought to be eventually overtaken by the more efficient {\it Pebble Accretion} process \citep{Lambrechts_2012,2017Johansen}. Pebble accretion continues until the solid core induces a pressure ``bump'' in the disc that halts further pebble inflow (\citealt{2006Paardekooper,2006Rice,2012Ayliffe}, later referred to as \emph{pebble isolation}; \citealt{Ataiee_2018}; \citealt{Bitsch_2018}). Since both stages involve the buildup of the planet's solid component, we collectively refer to them as {\it Solid Accretion}. After solid accretion halts, {\it Gas Accretion} can commence, whose rate depends on the effective cross-section of the planet and local disc conditions, and continues untils a gap opens in the planet's orbit or the disc becomes depleted of gas \citep{2010dangelo,drazkowska2023planetformationtheoryera}. 

In the following sections, we will estimate the effectiveness of each of these stages in the environments of post-AGB discs.

\subsection{The Dust Coagulation Phase in Post-AGB Discs}
\label{sec:grain-growth}

\begin{figure}[t]
\centering
\includegraphics[width=\linewidth]{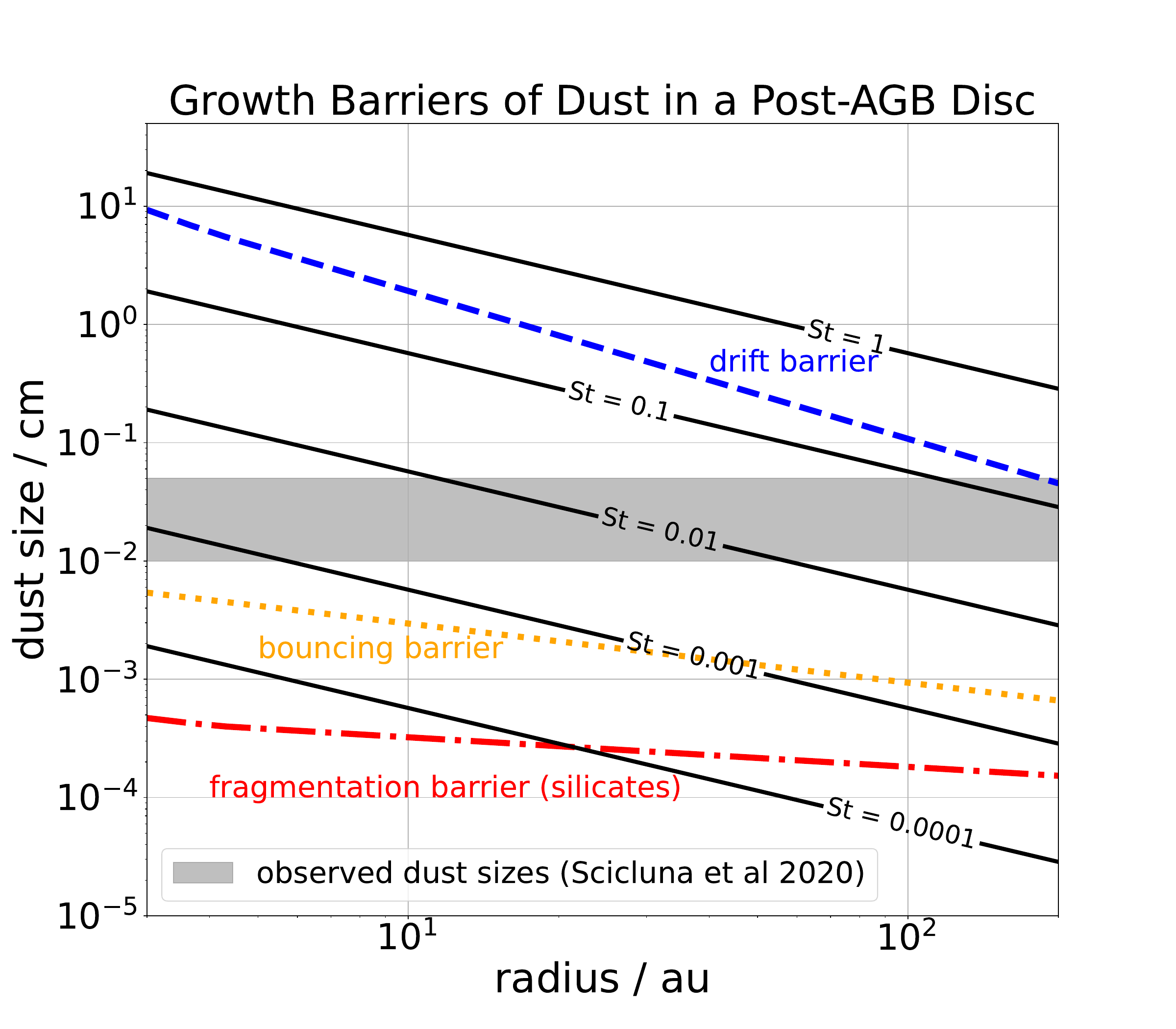}
\caption{Dust grain size vs. disc's radius for different Stokes number (black solid lines) for a representative post-AGB disc ($n=1$, $M_{\rm cnt}=1.5$M$_\odot$, $R_{\rm star} = 100$R$_\odot$, $T_{\rm star} = 5000 K$, $M_{\rm disc} = 0.02$M$_\odot$, $R_{\rm in}=3$au, and $R_{\rm out}=200$au) and dust-to-gas ratio of $0.01$. The gray shaded area indicates the maximum grain sizes observed in post-AGB discs by \citet{Scicluna_2020}, which are not spatially resolved. Upper growth limits set by the drift (blue dashed line), fragmentation (red dot–dash line; only for silicates), and bouncing (orange dot line) barriers are shown.
}
\label{fig:grains_pagb}
\end{figure}

In this section we determine whether the dust coagulation phase can lead to the grain sizes observed in post-AGB discs \citep{Scicluna_2020} or in any case create large enough grains to move on to the next stage of planet formation. Figure~\ref{fig:grains_pagb} shows dust grain size vs. radius, with a grey band indicating the upper end of the grain sizes we need to account for in post-AGB discs. The black curves are contours of Stokes number $S_{\rm t}$\footnote{To avoid confusion we stress that $S_{\rm t}$ is used in other fields to denote the Strouhal number while the Stokes number is denoted by Stk, however since in the planet formation community the Stokes number is usually denoted by $S_{\rm t}$ we will adopt this notation.}: large values mean that grains are not as easily slowed down in the disc's gaseous flow. The yellow, blue and red lines indicate the barriers to growth in this regime. Details of how these barriers are calculated in the case of post-AGB discs are reported in \ref{ap:barriers}. 

As can be seen in Figure~\ref{fig:grains_pagb}, the main restriction for grains to grow to the observed sizes are the fragmentation and bouncing barriers, both of which depend on the poorly constrained dust composition. 
The fragmentation barrier depends on the assumed composition, on which in turn depends the threshold velocity for fragmentation. In Figure~\ref{fig:grains_pagb} we plot the fragmentation curve for silicates
\citep{2010guttler_zoo}; for all other compositions this barrier shifts to grain sizes larger than observed in post-AGB discs.
For the case of the bouncing barrier, the growth restriction may be alleviated if grains are porous rather than compact, since the bouncing barrier in Figure~\ref{fig:grains_pagb} is computed for compact aggregates, or if relative collision velocities are sufficiently high \citep[][for more information see \ref{ap:barriers}]{Loesche_2016,Birnstiel_2016}  
 
Next we estimate the timescale needed for dust coagulation up to mm-sizes in post-AGB discs. Estimations for post-AGB disc lifetimes are around $10^{4-5}$ years \citep{Bujarrabal_2017,martin2025modellingimpactcircumbinarydisk}, a time that we set as an upper limit for the entire growth sequence to form a planet. Many post-AGB discs have inferred millimeter grain sizes, so the expectation is that grain growth occurs early in the disc's life, possibly within $<1000$ years \citep{Scicluna_2020}. Here we therefore determine whether these timescales are compatible with YSO dust coagulation theory, but using the physical characteristics of post-AGB discs.

Many studies simulate grain growth using the Smoluchowski equation \citep[e.g.,][]{2010Birnstiel,Dr_kowska_2014}. Due to its complexity, several works instead use the analytical framework of \citet{1997StepinskiValageas}; e.g., \citet{Laibe_2008,2024Michoulier}, which assumes a narrowly peaked grain size distribution and perfect sticking upon collision. Both approaches neglect alternative collisional outcomes\footnote{\citet{2024Michoulier} have coupled some of the destructive outcomes from experiments with the method of \citet{1997StepinskiValageas} in their simulations. However, doing so requires details such as the porosity of the dust particles among others, which are currently not available for post-AGB discs and is out of the scope of our study.}.

\citet{1997StepinskiValageas} track the evolution of a characteristic grain size, $a(r,t)$, at each disc radius. Assuming perfect sticking, they derive an equation for the rate of particle size growth over time
\begin{equation}
    \label{eq:grain_growth_rate1}
    \frac{da}{dt}=\frac{\rho_{\rm d}(r)}{\rho_{\rm s}}v_{\rm rel},
\end{equation}
where $\rho_{\rm d}(r)$ is the mass of solids per volume as a function of radial position (usually taken to be $0.01\rho_{\rm g}(r)$ with $\rho_{\rm g}$ being the volume density of the gas component of the disc), $\rho_{\rm s}$ is the internal grain density (the dust aggregate density is assumed to be 1.6 g~cm$^{-3}$) and $v_{\rm rel}$ is the mean relative velocity between the solid particles
\begin{equation}
    \label{eq:v_rel}
    v_{\rm rel}=\sqrt{2}v_t\frac{\sqrt{S_{\rm c}-1}}{S_{\rm c}},
\end{equation}
where $v_t = c_s\sqrt{2^{0.5}R_{\rm o}\ \alpha}$ is the turbulent velocity, $\alpha$ the Shakura-Sunyaev viscosity parameter, $R_{\rm o}=3$ the Rossby number for turbulent motions \citep{2024Michoulier} and $S_{\rm c}$ is the Schmidt number of the flow, which measures the grain’s coupling to the micro-scale turbulent motions
\begin{equation}
\label{eq:schmidt_number}
    S_{\rm c}=(1+ S_{\rm t})\sqrt{1+\left(\frac{\bar{v}}{v_t}\right)^2},
\end{equation}
where $\bar{v}\approx c_s^2/v_{K}$ is the mean relative velocity between gas and dust not to be confused with $v_{\rm rel}$ (it can be shown $\bar{v}\approx c_s^2/v_{K}$, see \ref{sec:drift}) and $S_{\rm t}$ is the Stokes number which is defined as \footnote{The definition of the Stokes number depends on the relation between the mean free path of a particle in a fluid and its size, and this equation here is valid for the Epstein regime where the particle is small enough and is close to the midplane \citep{2010Birnstiel}.}

\begin{equation}
\label{eq:stokes_def}
    S_{\rm t} = \frac{\pi}{2}\frac{a\rho_{\rm s}}{\Sigma_{\rm g}},
\end{equation}
where $\Sigma_{\rm g}$ is the gas surface density.
Substituting the relevant quantities in Equation \eqref{eq:grain_growth_rate1} we obtain
\begin{equation}
    \label{eq:grain_growth_rate}
    \frac{da}{dt}= \frac{\rho_{\rm d}(r)}{\rho_{\rm s}}\ \sqrt{2^{1.5}R_{\rm o}\ \alpha}c_{\rm s}\frac{\sqrt{S_{\rm c}-1}}{S_{\rm c}}.
\end{equation}
Integrating Equation \eqref{eq:grain_growth_rate} we find an expression for the time it takes to grow a grain from $a_{\rm min}$ to $a_{\rm max}$
\begin{equation}
    \label{eq:time for grain growth}
    t(r)=\frac{1}{\sqrt{2^{1.5}R_{\rm o}\ \alpha}\frac{\rho_{\rm d}(r)}{\rho_{\rm s}}c_{\rm s}}\int_{a_{\rm min}}^{a_{\rm max}}\frac{S_{\rm c}}{\sqrt{S_{\rm c}-1}}\ da,
\end{equation}
which is a function of the local conditions at the radial distance of the disc. We can numerically calculate this integral for a post-AGB disc using $\rho_{\rm d}(r) = 0.01 \rho_g(r) =0.01\ \Sigma(r) / \sqrt{2\pi}H(r)$.

\begin{figure}[t]
\centering
\includegraphics[width=\linewidth]{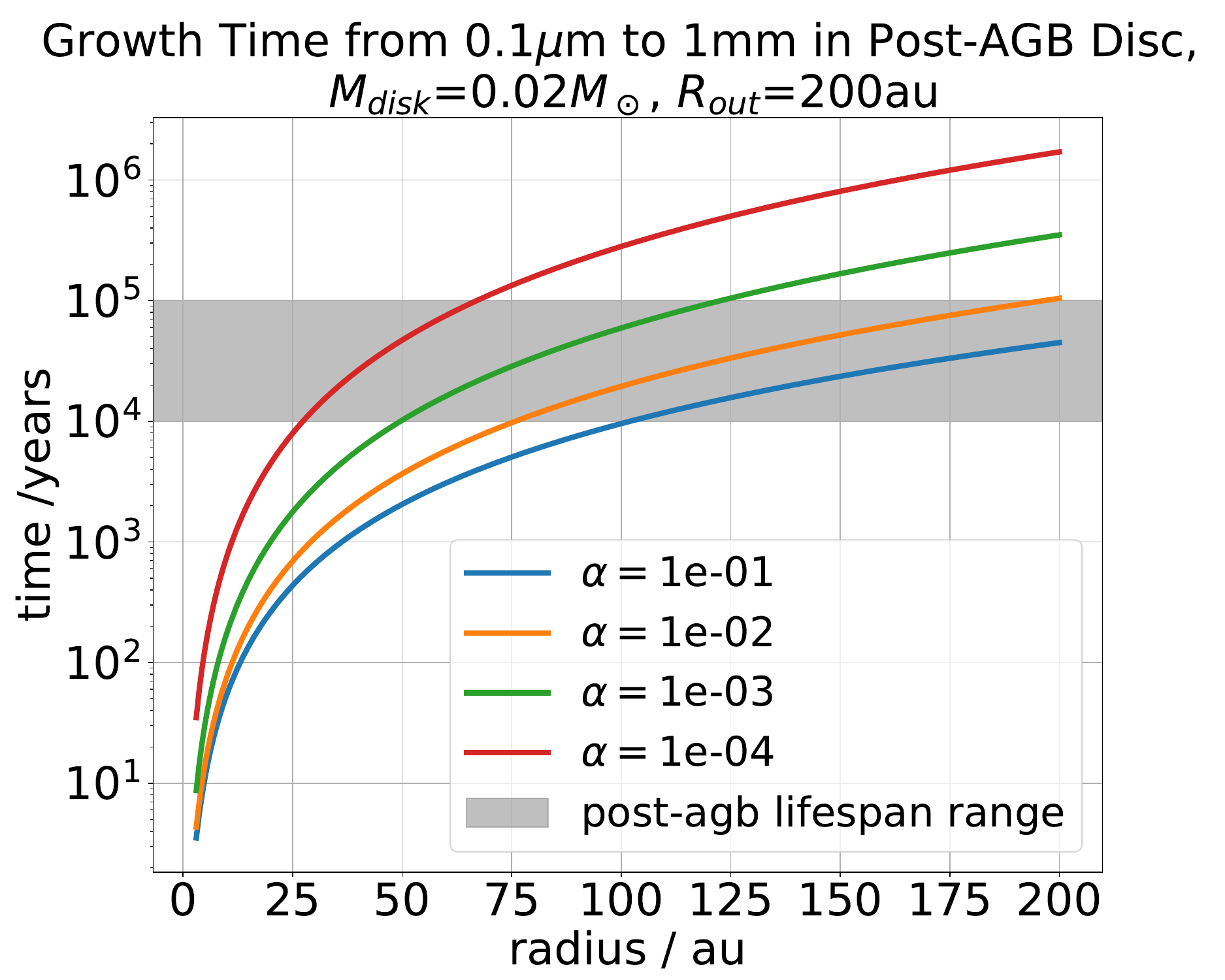}
\caption{Time taken by a grain to grow from $0.1\mu$m to 1~mm as a function of the post-AGB disc's radial location for different viscosity parameters $\alpha$, in a typical post-AGB disc (values similar to that of Figure \ref{fig:grains_pagb}). The grey shadowed area indicates the range of lifetimes estimated for post-AGB discs. For a sizable part of the inner disc, grain growth to $\sim1$~mm size is possible. Here it has been implicitly assumed that the grains are porous enough to bypass the bouncing barrier (see \ref{ap:barriers}, and Figure \ref{fig:grains_pagb}).}
\label{fig:time-mm}
\end{figure}

In Figure \ref{fig:time-mm} we plot the timescale to grow from 0.1~$\mu$m to 1~mm in a post-AGB disc with $M_{\rm disc}=0.02$~M$_{\odot}$, $R_{\rm out}=200$~au, and $n=1$. Grain growth is possible in the inner disc within short enough timescales compared to the disc's lifetime. The start value of $0.1\mu$m come from simulations of binary star interactions undergoing the common envelope phase \citep{bermúdezbustamante2024dustformationinteractionbinary,2024Siess} showing only year-long timescales for nucleation of dust. The range of $\alpha$ values shown spans from $10^{-4}$-$10^{-3}$, following \citet{2009Andrews} for YSOs, up to $10^{-2}$-$10^{-1}$ as inferred by \citet{Corporaal_2023} for the post-AGB disc IRAS08544-4431. 

The growth timescale increases with the outer disc's radius and with the assumed final grain size. For instance, in the post-AGB disc of Figures~\ref{fig:grains_pagb} and \ref{fig:time-mm} with $\alpha=0.001$, grains at 100~au reach $a_{\rm max}=1$~mm after $\sim0.06$~Myr, compared to $\sim0.03$~Myr and $\sim0.12$~Myr for $0.5$~mm and $2$~mm, respectively. We also estimated the growth timescales for discs irradiated by a solar-type star (Equation~\ref{eq:aspectratio}) to compare dust coagulation in post-AGB discs and PPDs around YSOs. While the timescale does vary between the two cases, the difference is not systematic: depending on the assumed value of $\alpha$, it can be either longer or shorter in PPDs around YSOs. In all cases, however, the order of magnitude remains comparable.

\subsection{Planetesimal Formation: Gravitational Collapse of Dust Particles and the Streaming Instability}
\label{sec:streaming}

The radial drift barrier ensures that dust particles cannot  coagulate beyond a certain size. Further growth requires the formation of an over-dense region of dust particles that can collapse gravitationally into larger bodies, at which point radial drift becomes negligible. 

Several mechanisms have been proposed that can concentrate dust particles, including particle trapping in dead zones, turbulence-induced vortices, and pressure bumps (see \citealt{Cuzzi_2001,Birnstiel_2016,drazkowska2023planetformationtheoryera}). Any mechanism capable of producing gravitationally bound solids that produce planetesimal-sized objects can fit within the core accretion framework to help explain further growth. 

For the purpose of this paper, we assume that the ``streaming instability'' \citep[SI;][]{Youdin_2005} allows growth of solid  bodies past the drift barrier to an adopted planetesimal mass value, $M_{\rm psml}$ (Figure~\ref{fig:growthpath}), which serves as the starting point for subsequent growth. We do so in the understanding that this mechanism is not itself well understood and whether it  operates in PPDs around YSOs is still the subject of intense scrutiny (see \citealt{johansen2015growthasteroidsplanetaryembryos,2016Simon,Abod_2019,2019Krapp}). 

The SI aims to explain a concentration of solid particles through the mutual drag between nearly Keplerian dust and sub-Keplerian gas. As a result, the local dust density becomes high enough that the gas is accelerated toward Keplerian speed, reducing the headwind and further enhancing particle concentration. Then, if the solid clusters exceed the Roche density, they can gravitationally collapse into $\sim100$~km planetesimals \citep{Youdin_2005,2017Carrera}. 

\begin{figure}[hbt!]
\centering
\includegraphics[width=0.9\linewidth]{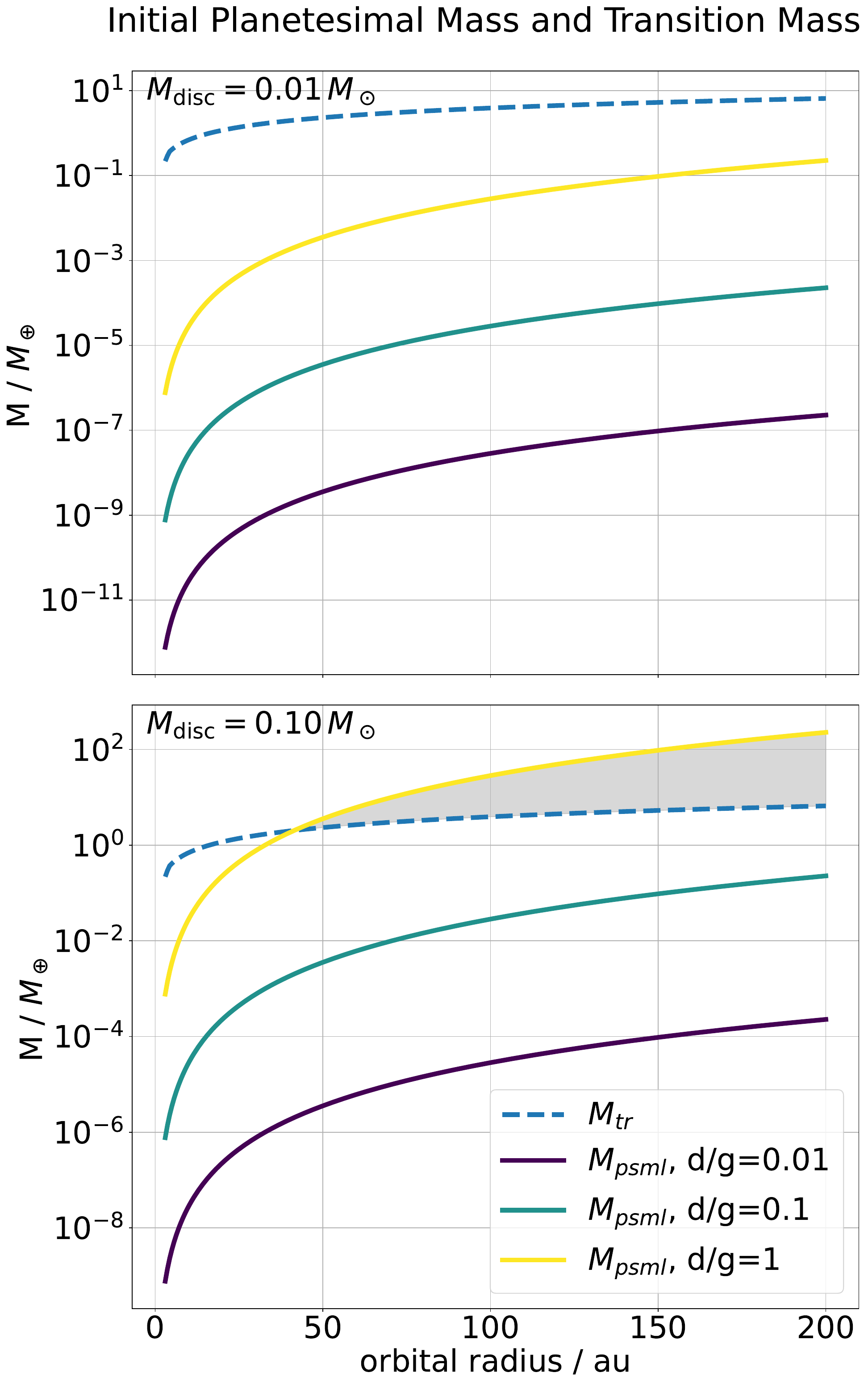}
\caption{The Initial planetesimal masses (solid lines) that can form from gravitational collapse (via SI) as a function of radius are shown here, for three different local dust-to-gas ratios (denoted by d/g) and two different disc masses. Additionally, the transition mass $M_{\rm tr}$ between the Bondi and Hill regimes of pebble accretion for each disc mass is also depicted with dashed lines (see Section~\ref{sec:pebble_accretion}, Equation~\eqref{eq:transitionmass}). Properties of the central binary are taken to be similar to the properties of AC Her, as described in \ref{ap:obs_features_pagb}. It can be seen that for the 0.1~M$_\odot$ disc mass, if the dust-to-gas ratio is equal to 1, the products of SI bypass the transition mass $M_{\rm tr}$ for r>42~au.}
\label{fig:SI}
\end{figure}

Gravitational collapse requires the mass of the solids' concentration to exceed the gravitational mass $M_G$, given by \citep{Abod_2019}
\begin{equation}
    \label{eq:gravitational_mass}
     M_{\rm G}=\frac{4\pi^5G^2\Sigma_d(r)^3}{\Omega_K(r)^4}.
\end{equation}
Planetesimals formed through gravitational collapse instigated by the SI will follow a mass distribution with a characteristic planetesimal mass above which the number of planetesimals steeply declines (the high-mass tail of SI planetesimal mass distributions remains uncertain). Following figure 8 in \citet{Abod_2019}, for a post-AGB disc with $\Pi\sim0.15$ we take the initial planetesimal mass to be $M_{\rm psml}=0.6M_G$ as the representative mass produced by these collapse processes (Figure~\ref{fig:growthpath}) and use Equation~\eqref{eq:gravitational_mass} to plot the initial planetesimal mass as a function of orbital distance in Figure~\ref{fig:SI} for two different disc masses. We find that for a disc with a mass of $0.1$~M$_\odot$, the products of the SI can reach a mass larger than a threshold value, $M_{\rm tr} \sim 10$~M$_\oplus$, above which pebble accretion becomes significant enough to be a dominant mechanism for further growth (as explained in Section~\ref{sec:pebble_accretion}).

Next we ask if the process is fast enough to operate within the short disc's lifetime. In this regard, notwithstanding the fact that the timescale for the SI depends on an assumed dust size distributions \citep{2019Krapp}, under ideal conditions growth timescales can be as short as $\lesssim100/\Omega_K$ \citep{Youdin_2005} which for a disc with a central mass of $1.5$~M$_\odot$ at $r=50$au is $\sim5000$ years; we therefore assume that the timescale over which the SI promotes the formation of planetesimals is short enough to fit within the observed timescale. 

Several conditions identified in simulations as favorable for the SI are likely met in post-AGB discs. First, for particles of similar size, the SI is typically triggered for Stokes numbers $0.01 \lesssim S_{\rm t} \lesssim 3$, midplane dust-to-gas ratios $\gtrsim 1$, and supersolar metallicities \citep[]{2010Bai_a,Krijt_2016_panoptic}. In Figure~\ref{fig:grains_pagb} we showed that beyond $r \gtrsim 10$ au, grain sizes correspond to $S_{\rm t} \gtrsim 0.01$ , satisfying the Stokes number requirement. Second, enhanced dust-to-gas ratios also make the SI more likely \citep{2010Bai_a}, which is inferred from observations of the post-AGB disc AC~Her with dust-to-gas ratios potentially exceeding 0.1\footnote{\citet{Hillen_2015} reported dust-to-gas ratios exceeding 0.1, though these measurements have been disputed in other studies \citep[e.g.,][]{Gallardo_Cava_2021,martin2023acherevidencepolar}.}. Third, the large aspect ratios of post-AGB discs result in a substantial difference between the gas and Keplerian velocities. Gas in a pressure-supported disc orbits at a sub-Keplerian velocity \citep{Youdin_2005}
\begin{equation}
\label{eq:subkep}
    v=\sqrt{v_{\rm K}^2+\frac{r}{\rho_{\rm G}}\frac{dP}{dr}}=v_{\rm K}\sqrt{1-2\eta}, 
\end{equation}
with $P$ being pressure, where $\eta$ is defined as
\begin{equation}
\label{eq:eta}
    \eta=-\frac{r}{2\rho v_K^2}\frac{dP}{dr} =    -\frac{p-2n-3}{4}\left(\frac{H}{r}\right)^2.
\end{equation}
where the second expression assumes an isothermal disc, a power-law surface density $\Sigma\propto r^{-n}$, and a temperature profile of $T\propto r^{p}$. Since the SI is driven by relative motion between solids and gas, a larger $\eta$ corresponds to a greater gas--particle velocity difference. Thus, the larger aspect ratios of post-AGB discs imply stronger relative drift than in PPDs around YSOs, potentially favouring the onset of the SI.

A caveat is that \citet{Abod_2019} performed shearing-box SI simulations varying the dimensionless pressure-gradient parameter $\Pi=\eta v_k/c_s\propto (H/r)^{-1}$ and found that larger $\Pi$ reduces planetesimal formation on the timescales of their simulations. However, they concluded that their simulation times were too short for firm conclusions, because  overdense filaments persist at high $\Pi$ and may still continue to form planetesimals beyond their simulated time. Since $\Pi$ is systematically larger in post-AGB discs than in PPDs around YSOs (due to higher $H/r$), the SI may operate in a regime where filament formation is robust but slower, hence planetesimal formation could be delayed rather than suppressed. 

\subsection{Solid Accretion in Post-AGB Discs}
\label{sec:pebble_accretion}

The next viable stage of planet growth in post-AGB discs only takes place for the more massive discs we have considered in Section~\ref{sec:streaming} ($\sim0.1$M$_\odot$), where, for higher local dust-to-gas ratios of approximately unity, bodies have grown via the SI to masses larger than the transition mass, $M_{\rm tr}$. 

Above the transition mass pebbles can accrete onto the planetesimal via ``Hill regime pebble accretion''. Pebbles are defined as solid particles with $0.001\text{-}0.01 < S_{\rm t} < 1$, implying that they are less coupled to the disc gas than  smaller dust particles. With this definition, pebble sizes vary throughout the disc. As shown in Figure~\ref{fig:grains_pagb}, the observed grains  in post-AGB discs correspond to $0.001 < S_{\rm t} < 1$ over most of the radial extent, implying that they are predominantly in the pebble regime. 

The Hill regime, refers instead to when accretion onto the planetesimal is for pebbles inside the body's Hill radius. This happens when the planetesimal's mass grows past the point when its Bondi radius becomes comparable to its Hill radius and defines the transition mass, $M_{\rm tr}$, drawn in Figure \ref{fig:SI} (dashed lines) and defined by \citet{2017Johansen}
\begin{equation}
    \label{eq:transitionmass}
    M_{\rm tr}=\sqrt{\frac{1}{3}}\frac{\Delta v^3}{G\Omega_{\rm K}},
\end{equation}
where $\Delta v$ is the relative velocity between the pebbles and the solid planetary core which is equal to \citep{1977Weidenschilling,1986Nakagawa}
\begin{equation}
    \label{eq:vrel}
\Delta v= \frac{\sqrt{\rm 4S_{\rm t}^2+1}}{\rm S_{\rm t}^2+1}\eta v_{\rm K} \approx \eta v_{\rm K},
\end{equation}
for $S_{\rm t}\ll$1, typical of post-AGB discs, which depends only on the central binary and not on the disc's properties. Using Equation \eqref{eq:eta} for $\eta$ and rearranging Equation \eqref{eq:transitionmass} we find
\begin{equation}
    \label{eq:transitionmasshr}
    M_{\rm tr}=  \left(-\frac{p-2n-3}{4}\right)^3\frac{rv_K^2}{\sqrt{3}G}\left(\frac{H}{r}\right)^6.
\end{equation}

In Figure \ref{fig:SI} the transition mass $M_{\rm tr}$ varies between $0.1-10$~M$_\oplus$ in post-AGB discs, which is $\sim2$ orders of magnitude larger than the transition mass in PPDs around YSOs, due to the higher aspect ratios of these discs. Hill regime pebble accretion can only start once bodies grow past this significantly higher transition mass.

In Figure \ref{fig:SI} we also see that for a 0.1~M$_\odot$, post-AGB disc and a local dust-to-gas ratio of unity, the planetesimal mass created from gravitational collapse from SI ($M_{\rm psml}$) can be larger than $M_{\rm tr}$ for radii of $r\gtrsim42$~au. Under these conditions planetesimals created by gravitational collapse can immediately start accreting pebbles within their Hill radius. In Figure~\ref{fig:SI} the disc mass values of 0.1 and 0.01~\msun{} were meant to show representative cases for which $M_{\rm tr}$ can or cannot be reached. We note that within that range, even a post-AGB disc mass as low as 0.05~\msun{} could promote a planetesimal mass growth exceeding the transition value.

The timescale for Hill regime pebble accretion is \citep{Lambrechts_2012}
\begin{equation}
    \label{eq:hill_time}
    \Delta t_{\rm H}= \int_{M_{\rm tr}}^{M_{\rm f}}\dot{M_{\rm H}^{-1}}dm\approx \frac{3^{5/3}\Omega_{\rm K}(r)^{1/3}}{2G^{2/3} \Sigma_d(r)}M_{\rm f}^{1/3},
\end{equation}
which is valid when the starting mass (set to $M_{\rm tr}$) is much smaller than the final mass. $M_{\rm tr} \ll M_{\rm f}$. If this is valid, the timescale isn't sensitive to $M_{\rm tr}$ and mainly varies with $M_{\rm f}$. We set the final mass to be the ``pebble isolation mass'', $M_f=M_{\rm iso}$ , the mass at which the solid core induces a pressure maximum outside of its orbit in the disc that halts further pebble inflow. 

\begin{figure}[hbt!]
\centering
\includegraphics[width=0.9\linewidth]{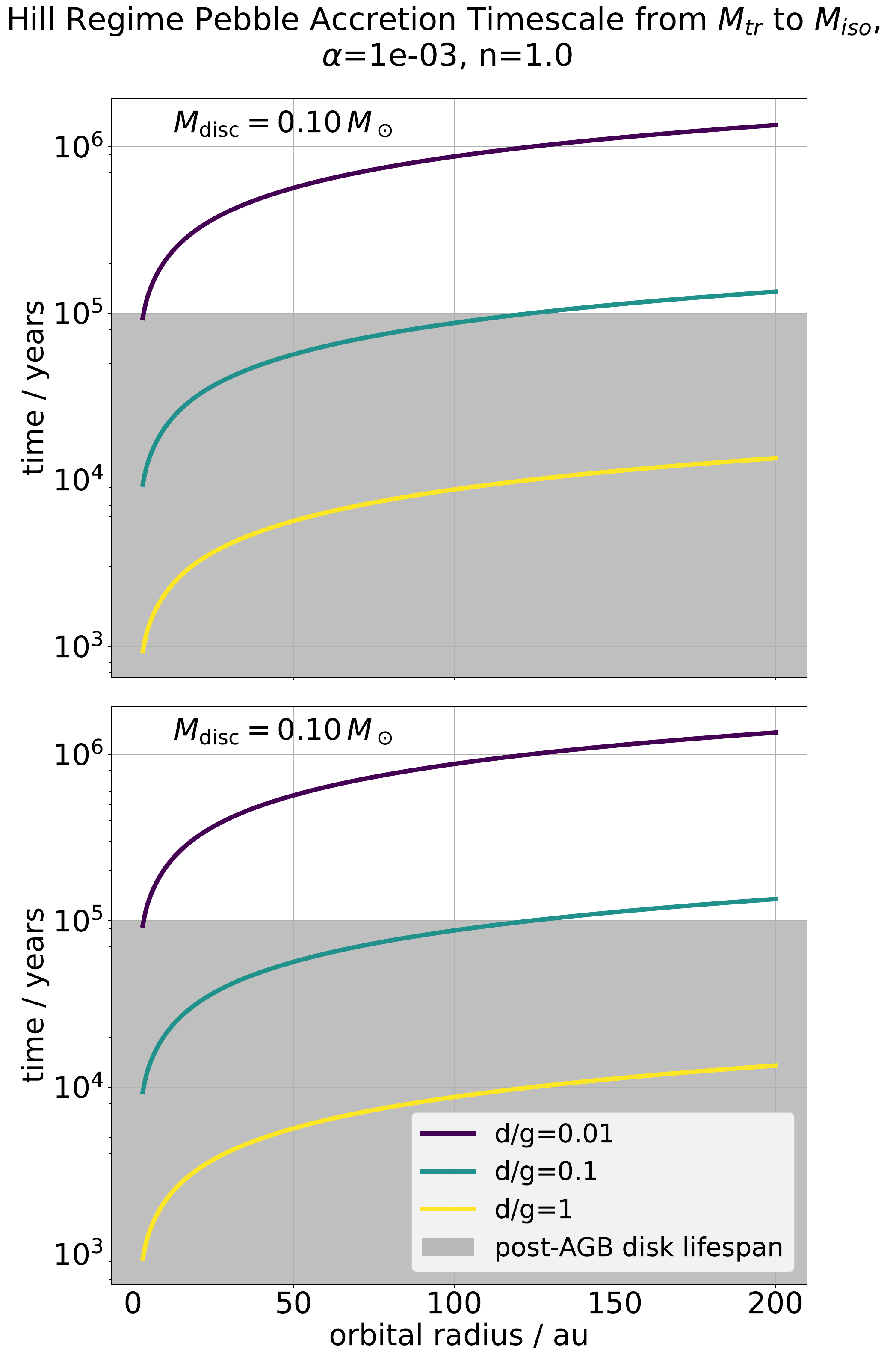}
\caption{Hill Timescales for planetesimal growth as a function of orbital radius in post-AGB discs for three different local dust-to-gas ratios (denoted by d/g) and two different disc masses, starting from $M_{\rm tr}$ up to $M_{\rm iso}$. Properties of the central binary are taken to be similar to the properties of AC Her, as described in \ref{ap:obs_features_pagb} (similar to Figure \ref{fig:SI}).}
\label{fig:timescales_pebble}
\end{figure}

In Figure \ref{fig:timescales_pebble} (bottom panel) we show the time it takes to grow to the pebble isolation mass for different dust-to-gas ratios over the disc's extent. For the case of $M_{\rm disc}=0.1$M$_\odot$ we see that Hill accretion allows growth to $M_{\rm iso}$ within $\sim 10^5$ years over parts of the disc, particularly for enhanced dust-to-gas ratios. Hence, if the initial planetesimal mass is larger than the transition mass, efficient growth within post-AGB disc lifetimes is possible\footnote{We adopt $\alpha = 10^{-3}$ to estimate the pebble isolation mass (Equation \eqref{eq:pebble_isolation_mass}), which determines the final mass. This choice does not significantly affect the growth timescale.}. 

Additionally, we find that {\it the pebble isolation masses in post-AGB discs ($\sim 1$-$10$~M$_{\rm J}$) are 1--2 orders of magnitude higher than those typically expected in PPDs around YSOs}, primarily due to the larger aspect ratios characteristic of these discs. This implies that, in principle, \textit{rocky planets with masses up to and comparable to Jupiter} could form in these environments, neglecting other constraints such as the total solid mass budget of the disc, as well as the stability of such a solid body with higher masses. 

\begin{figure}[hbt!]
\centering
\includegraphics[width=0.9\linewidth]{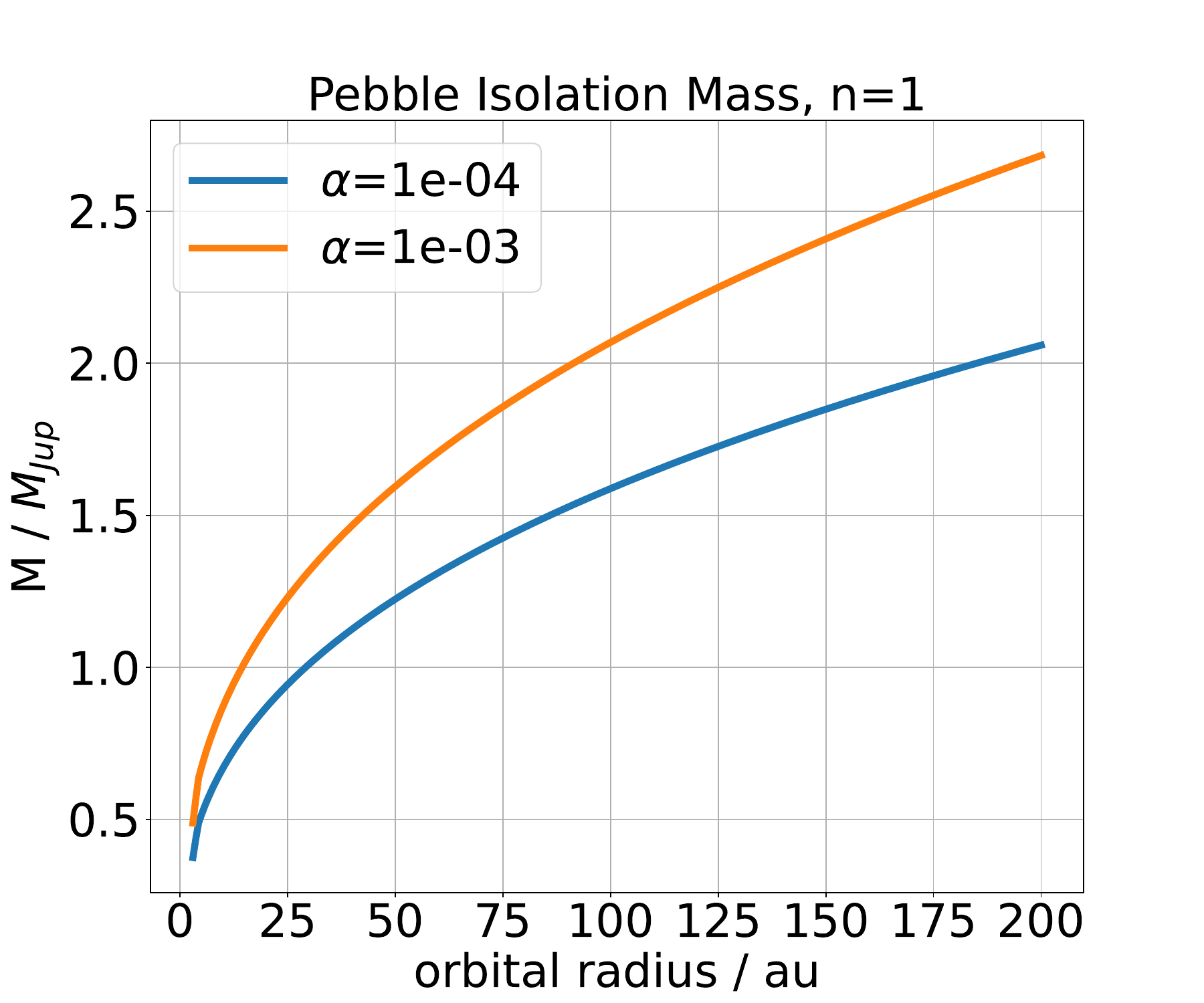}
\caption{Pebble isolation mass as a function of orbital radius in a post-AGB disc for 3 different values of $\alpha$. Properties of the central binary are taken to be similar to the properties of AC Her, as described in \ref{ap:obs_features_pagb}. We note that varying the disc mass does not affect the pebble isolation mass, thus there may not be enough solid mass in post-AGB discs for a planet to reach these isolation masses (e.g. setting $\alpha=0.01$ results in isolation masses higher than the solid disc budget of AC Her).}
\label{fig:pebble_isolation_mass}
\end{figure}

In YSOs, pebble isolation masses are typically expected to be $\sim10$–$20,M_\oplus$, broadly consistent with theoretical models and Solar System constraints, with estimated core masses of $\sim7$–$25,M_\oplus$ for Jupiter and $\sim15$–$18,M_\oplus$ for Saturn, although some rocky exoplanets such as TOI-849b may exceed this range \citep{2017Alibert,drazkowska2023planetformationtheoryera,Armstrong_2020}. Several expressions exist for the pebble isolation mass $M_{\rm iso}$ \citep{Lambrechts_2014,Ataiee_2018,Bitsch_2018}. In this work, we adopt the relation found by \citet{Bitsch_2018} which is

\begin{align}
    \label{eq:pebble_isolation_mass}
    M_{\rm iso}&= 25M_\oplus \left(\frac{H/r}{0.05}\right)^3 \notag \\
    &\times\left(0.34\left(\frac{\log{10^{-3}}}{\log{\alpha}}\right)^4+0.66\right)\left(1-\frac{\frac{\partial \ln P}{\partial \ln r}+2.5}{6}\right),
\end{align}
thus $M_{\rm iso}$ is most sensitive to the aspect ratio. Using Equation \eqref{eq:eta} to derive $\partial \ln P/\partial \ln r$, we can estimate the pebble isolation mass in post-AGB discs (see Figure \ref{fig:pebble_isolation_mass}). 

For the low-mass post-AGB discs where the planetesimal masses are well below the transition value (Figure~\ref{fig:SI}, top panel), growth via Hill regime pebble accretion is not initially possible. In such a case the body can grow either through ``planetesimal accretion'', i.e., collisions of planetesimals of similar size \citep{Lambrechts_2012}, or through ``Bondi regime pebble accretion''. Neither mechanisms afford growth within reasonable timescales, but of the two the planetesimal accretion is the fastest ($\sim 2 \times 10^9$~yr; \citealt{Goldreich_2004,Lambrechts_2012,Dodson_Robinson_2009}).

Next we show that, if sufficient disc material and time remain, gas accretion onto planetesimals that have grown to the pebble isolation mass can form super-Jupiter type planets.

\subsection{Gas Accretion and Migration}
\label{sec:gas_accretion}

Gaps and other asymmetries have been observed in post-AGB discs (see \ref{ap:obs_features_pagb}).  If they are caused by a planet, that planet must have grown large enough to satisfy the gap-opening criterion. In order for a planet to grow to this ``gap opening mass",  growth beyond the pebble isolation mass is required, which can happen via gas accretion \citep{1980Mizuno,2006Rafikov,Lambrechts_2012}. A planet with a mass commensurate with the pebble isolation mass interacts with the disc and can undergo migration.

To calculate planet mass growth due to gas accretion and orbital decay due to migration, we solve the following coupled differential equations
\begin{equation}
\label{eq:dmdtmigr}
 \frac{dM_P}{dt}= \frac{\Sigma(r)\pi R_{\rm eff}^2 c_s}{\sqrt{2\pi}r},
\end{equation}
\begin{equation}
\label{eq:drdtmigr}
    \frac{da}{dt}=\frac{2C_I\Sigma(r) r^3 \Omega(r) \left(\frac{r}{H}\right)^2\left(\frac{M_{\rm p} }{M_{\rm cnt}}\right)^2}{M_{\rm p} },
\end{equation}
where $a$ denotes the orbital separation of the growing planet which for the assumed circular orbit is the same as $r$; $M_P$ is the planet's mass, $R_{\rm eff}$ is the cross-section for gas accretion given by the smaller of the Hill and Bondi radii\footnote{This is different from the Bondi radius mentioned in Section~\ref{sec:pebble_accretion}.} (here $R_B=GM_{\rm P}/{c_s^2}$), and $C_I$ is a dimensionless constant of order unity \citep{2010dangelo,paardekooper2022planetdiskinteractions}. Equation~\eqref{eq:drdtmigr} can be derived by equating the angular momentum of the planet with the torque due to Type~I migration \citep{2010dangelo}, and  Equation~\eqref{eq:dmdtmigr} can be derived using the rate of disc gas swept by the accreting cross section of the planet. We use a fourth-order Runge-Kutta numerical scheme to integrate these equations and impose a time limit of $10^5$ years (typical lifetimes of post-AGB discs).

We halt the evolution once the planet is capable of opening a gap in the disc. To ensure this, we impose two conditions. First, we check whether the planet is sufficiently massive to open a gap (Equation~\eqref{eq:kanagawa}; in \ref{ap:obs_features_pagb}). Second, we verify that the planet has enough time to actually form the gap. For this, we compare the gap-opening timescale, $t_{\rm gap}$, with the planet crossing time. The gap-opening timescale is given by \citet{1986lin}
\begin{equation}
\label{eq:tgap}
t_{\rm gap}\approx\left(\frac{M_*}{M_{\rm p}}\right)^2\left(\frac{H}{r}\right)^5\frac{1}{\Omega(r)}.
\end{equation}
The planet crossing time is approximated as $t_{\rm cross}\approx\frac{\Delta_{\rm Gap}}{da/dt}$, assuming a gap width of $\sim\Delta_{\rm Gap}$ \citep{1986lin} (Equation~\ref{eq:gap_width}). We require that $t_{\rm gap} < t_{\rm cross}$ so that the planet can open a gap before migrating across it. Only when both conditions are satisfied do we stop the simulation \footnote{By doing so, we are neglecting the effects of the slower type II migration; \citep{2010dangelo}.}.

\begin{figure*}[t]
\centering
\includegraphics[width=0.8\textwidth]{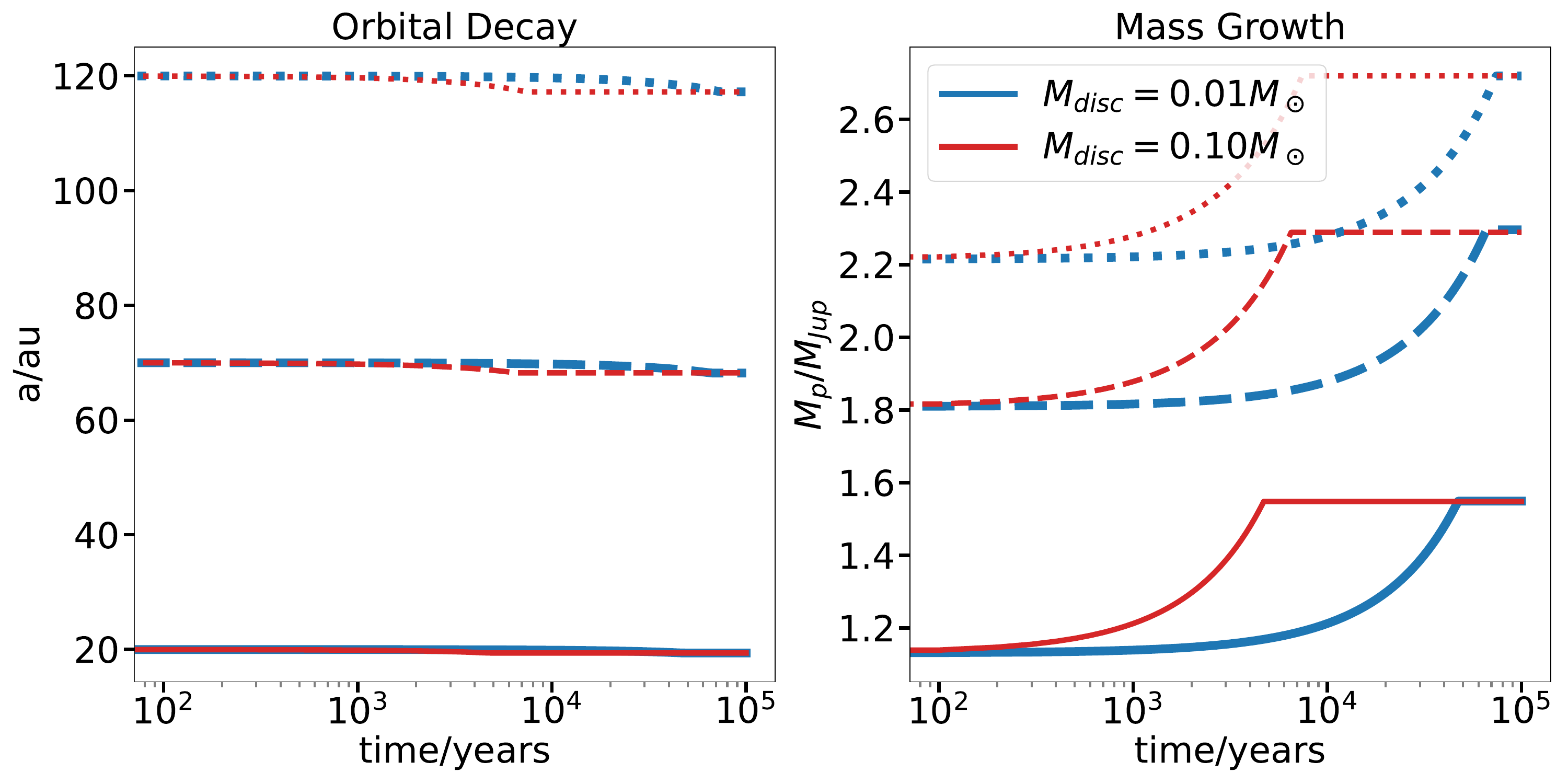}

\caption{ Orbital decay (left panel) and mass growth (right panel) for planets placed initially at 20 (solid line),70 (dashed line)and 120 au (dotted line) with initial masses of $M_{\rm P,init}=M_{\rm iso}$, embedded in 2 different post-AGB discs with masses of 0.01~M$_\odot$ (blue lines) and 0.1~M$_\odot$ (red line). Properties of the central binary are taken to be similar to the properties of AC~Her (see \ref{ap:obs_features_pagb}) with $\alpha=0.001$, $n=1$, $r_{\rm in}=3$~au  and $r_{\rm out}=200$~au. The planets that plateau have opened a gap within $10^5$ years .}
\label{fig:gasaccretion}
\end{figure*}

Figure \ref{fig:gasaccretion} shows the results of our simulations for two different post-AGB disc masses, assuming a common outer radius of $r_{\rm out}=200$~au, $r_{\rm in}=3$~au, $n=1$, $\alpha=0.001$ and stellar parameters corresponding to AC Her's central binary. For each disc, we simulate gas accretion for solid cores initially located at $a=20$, $70$, and $120$ au, with starting masses equal to the pebble isolation mass at those radii (Equation~\ref{eq:pebble_isolation_mass}). For both disc masses of $0.01$ and $0.1$M$_\odot$ growth is sufficient to open a gap within the lifetime of the post-AGB disc and for the higher mass, it takes less than $10^4$ years, however with the adopted value of $\alpha$, orbital decay is negligible in all scenarios. The gap-opening mass achieved is a function of the viscosity parameter, $\alpha$, as shown in Table \ref{tab:gasaccretion}. It can be seen that for higher $\alpha$ the orbital decay is more notable, as well as the final mass being larger.

\begin{table}[t]
\resizebox{0.8\textwidth}{!}{%

\begin{threeparttable}
\caption{Effect of the viscosity parameter, $\alpha$, on orbital migration (between $a_{\rm init}$ and $a_{\rm final}$) accomplished within the gap opening time, $t_{\rm gap}$ and the planet's mass ($M_{\rm p, final}$) achieved in that time, for a $0.1$M$_\odot$ disc. All planets are assumed to begin at the pebble isolation mass, $M_{\rm iso}$. For $\alpha=10^{-4}$, all of the planets already satisfy the gap-opening condition, hence they do not migrate at all.}

\label{tab:gasaccretion}
\begin{tabular}{cclcl}
\toprule

   $\alpha$& $a_{\rm init}$&$a_{\rm final}$&$t_{\rm gap}$  &$M_{\rm p, final}$\\
   
& (au)&(au)&(yrs)  &$(M_{\rm J}$)\\
\midrule

  $10^{-2}$& $20$&$16$& $1\times10^4$&$4.6$\\
   $10^{-3}$& $20$&$19$& $4\times10^3$&$1.6$\\
 $10^{-4}$& $20$&$20$& $0$&-\\
  $10^{-2}$&$70$&$58$&  $2\times10^4$&$7$\\
  $10^{-3}$& $70$&$68$& $6\times10^3$&$2.2$\\
  $10^{-4}$& $70$&$70$& $0$&-\\
   $10^{-2}$& $120$&$99$& $2\times10^4$&$8.1$\\
   $10^{-3}$& $120$&$117$& $8\times10^3$&$2.7$\\
   $10^{-4}$& $120$&$120$& $0$&-\\
  \bottomrule

\end{tabular}
\end{threeparttable}%
}
\end{table} 
The final planet / brown dwarf masses we reported in Table~\ref{tab:gasaccretion}, should be regarded as upper limits because if the accreted gas does not cool rapidly enough, it may become unbound; moreover we haven't factored out the timescales required for the prior stages of planet growth.

\section{Discussion and Summary}
\label{sec:summary}


Hints of planet formation have been observed in the discs around post-AGB binaries. Using the physical parameters of these discs, we have therefore carried out an approximate but quantitative analysis of planet formation, adopting the  core accretion  theoretical framework developed for PPDs around YSOs, in the understanding this framework still has several, very uncertain aspects.
Table~\ref{tab:comparisonsummary} lists the stages of core accretion and whether each stage is similar or not in post-AGB discs and PPDs around YSOs.

\begin{table*}[t]
\resizebox{\textwidth}{!}{%

\begin{threeparttable}
\caption{Comparison of the timescales and effectiveness of the different stages of core accretion between PPDs around YSOs and post-AGB discs.}

\label{tab:comparisonsummary}

\begin{tabular}{l l ll}
\toprule
Quantity & Metric &  Dependence on Aspect Ratio& Main result \\
\midrule
Dust growth & Timescale &Weak / non-trivial  & Comparable \\
Streaming instability & Timescale &  Uncertain& Uncertain \\
Streaming instability & Planetesimal mass produced &  nearly independent&Comparable \\
Pebble accretion (Hill transition) & Threshold mass &  $\propto(H/r)^6$&$\sim10^3$ times larger in post-AGB discs\\
Pebble accretion (Hill regime) & Timescale &  none& Comparable (for same final mass)\\
Pebble isolation & Isolation mass &  $\propto(H/r)^3$&$\sim10$--$10^2$ times larger in post-AGB discs\\
Gap opening & Threshold mass &  $\propto(H/r)^{2.5}$&$\sim10$--$10^2$ times larger in post-AGB discs \\
\bottomrule
\end{tabular}
\end{threeparttable}%
}
\end{table*} 

Our first conclusion is that for higher-mass post-AGB binary discs ($\sim$0.1~M$_\odot$), the timescales needed for the stages of core accretion to proceed from dust-sizes up to gap-opening masses may fit within the estimated lifetimes of post-AGB discs ($\sim 10^4 - 10^5$~yrs). The first phase of dust growth, from sub-micron to mm sizes, takes place in the inner half of the disc feasibly within $\sim 10^5$~yr, with efficiency increasing for higher disc viscosity and disc density. The observed mm-sized grains in post-AGB discs can be reproduced with theoretical dust-growth tracks.

For the second phase, dust concentrations can collapse into sufficiently large planetesimals and continue to grow to the pebble isolation mass (when pebbles stop accreting onto the nascent planet) within $\sim10^5$~years, provided the local dust-to-gas ratio is high and assuming that the streaming instability operates. 
If sufficient time remains, gas accretion can continue as long as disc gas is available, with the rate of orbital decay and mass growth strongly dependent on the viscosity parameter $\alpha$. 

Planet buildup may, however, not be fast enough in lower-mass discs ($\sim$0.01~M$_\odot$). While dust can still readily grow to mm sizes in these discs, the primary bottleneck is the difficulty of producing sufficiently large planetesimals through streaming instability-induced gravitational collapse of the dust layer to trigger rapid pebble accretion. \footnote{Nevertheless, planets may still form in these lower mass post-AGB discs if first-generation planets or other lower-mass bodies survive stellar evolution and binary interaction. If sufficiently large, these bodies could serve as embryos for rapid pebble accretion.}

We stress that the most uncertain step in core-accretion theories of planet formation, both in our application to post-AGB discs and in the YSO literature, is planetesimal formation.

Another critical aspect of planet formation in post-AGB discs, even in higher-mass discs ($\sim$0.1~M$_\odot$), appears to be the need for a locally enhanced dust-to-gas ratio.
We hypothesise that the streaming instability may be more readily triggered in post-AGB discs than in PPDs, as their larger aspect ratio results in larger pressure-supported deviations from Keplerian rotation and hence stronger dust–gas relative velocities, which may facilitate local dust enhancement. Verifying this hypothesis requires a dedicated study, outside the scope of this paper; observational constraints on local dust-to-gas enhancements would also be valuable.

Finally, we find that the much higher pebble isolation masses in post-AGB discs ($\sim M_{\rm Jup}$) compared with PPDs around YSOs ($\sim25M_\oplus$), arising from their larger $H/R$, make it possible in principle for \textit{rocky Jupiters} --- high-mass rocky planets --- to form in these discs. This suggests that planets formed in post-AGB discs may differ substantially from those formed in PPDs around YSOs, potentially providing an observational means of distinguishing between planet populations.

\paragraph{Acknowledgments}
AP acknowledges the financial support provided by the International Macquarie Research Excellence Scholarship (iMQRES) program, received throughout the duration of this research. DK, ODM acknowledge the support from the Australian Research Council Discovery Project DP240101150. 

\printendnotes
\bibliography{refs}
\appendix
\section{Observational Features of Post-AGB Discs Relevant to Planet Formation}
\label{ap:obs_features_pagb}

This appendix summarises the observational signatures of planet formation in post-AGB discs that have been reported in the literature and motivate the present study. We describe these signatures in detail below, with a summary provided in Table~\ref{tab:post-agb-list}.

\subsection{Disc Kinematics and Stability}

ALMA CO interferometric maps show that many post-AGB discs are long-lived, stable, and quasi-Keplerian \citep[e.g.,][]{Bujarrabal_2015,Bujarrabal_2016,Bujarrabal_2018}. More recently, \citet{Gallardo_Cava_2021,Gallardo_Cava_2023} have shown that post-AGB binaries can exhibit a range of disc outflow morphologies \citep[e.g., the red rectangle;][]{2013Bujarrabal}. For example, AC Herculis is strongly disc-dominated, whereas some systems have substantial molecular outflows \citep[e.g., 89 Herculis and R Scuti;][]{Gallardo_Cava_2021,Gallardo_Cava_2023}.
\subsection{ Substructures and Inner Cavities} 
\label{sec:substructure}
\citet{Andrych_2023,andrych2025spherezimpolinsightsdiscsevolved} and \citet{deprins2026vltipionierimagingpostagbbinaries} detected arcs, gaps, asymmetries, and other substructures in several post-AGB discs with {\sc SPHERE}/VLT and VLTI. While such substructures may indicate disc-planet interactions, alternative explanations include midplane shadows, disc winds, outflows, and jets. 

Inner cavities provide another potential diagnostic of disc evolution and planet formation. Mid-IR interferometric observations reveal inner cavities in several post-AGB discs \citep{Corporaal2023}. In otherwise unperturbed systems, such cavities can be explained either by the dust sublimation radius or by dynamical truncation due to disc–binary interactions. The latter arises from Lindblad resonances and produces a truncation radius $R_{\rm dyn}\sim2$–$5$ times the binary separation, depending on the aspect ratio $H/R$ at the inner rim, the eccentricity, and mass ratio of the system \citep{Artymowicz1994,Hirsh_2020}. Although mass can flow across this gap due to accretion, it will not be able to close the gap \citep{1996Artymowicz}.

Several post-AGB systems have been discovered with inner (dust) cavities 2.5--7.5 times larger than the dust sublimation radius \citep[][Table~\ref{tab:post-agb-list}]{Corporaal2023}. While other probable explanations have been given to explain wide inner gaps for PPDs around YSOs such as photoevaporation, grain growth, dead zones, etc. \citep{van_der_Marel_2023}, some of these mechanisms can be disregarded in post-AGB discs, e.g., photoevaporation is generally expected to be inefficient because post-AGB stars do not emit high-energy photons \citep{Kluska_2022}.

Using the truncation prescriptions of \citet{Hirsh_2020}, \citet{Anugu_2023} showed that even dynamical truncation cannot account for the large cavity of AC~Her. To assess whether binary-driven dynamical truncation can explain these large cavities more generally, we follow \citet{Anugu_2023} and use the results of \citet{Hirsh_2020} to estimate $R_{\rm dyn}$ for the remaining systems in \citet{Corporaal2023}. The results are listed in Table~\ref{tab:innercavities}. AC~Her remains the only system whose cavity is clearly inconsistent with both dust sublimation and dynamical truncation. For CT~Ori, AD~Aql, and ST~Pup, unknown orbital parameters prevent equivalent estimates. However, comparison of the inner cavity of AD~Aql, $\sim18$~au, with the orbital parameters of other post-AGB discs suggests that dynamical truncation may also be insufficient in this case. Further observations of AD~Aql are therefore needed to clarify the origin of its inner cavity.
AC~Her therefore provides a particularly interesting case for exploring whether an embedded companion could account for its large cavity. We estimate the planet mass required to carve such a cavity under the assumption that it is planet-induced.

\textsc{Planet Mass Needed to Carve Inner Cavity in AC~Her}: 

We adopt the established planet–disc gap opening framework developed for PPDs around YSOs. If the cavity is planet-induced, it can be interpreted as a gap whose inner region has already been cleared by sublimation or disc truncation. A gap forms when the planetary torque exceeds the opposing viscous and pressure torques \citep{1986lin}. Following \citet{Kanagawa_2018}, the minimum planet mass required to open a gap, $M_{\rm p,gap}$, is given by

\begin{equation}
\label{eq:kanagawa}
    M_{\rm p,gap}=8\times10^{-5}\times M_{\rm cnt}\left( \frac{\alpha}{10^{-3}} \right)^{0.5}\left( \frac{\frac{H}{r}}{0.05} \right)^{2.5}
\end{equation}
where $\alpha$ is the Shakura-Sunyaev viscosity parameter \citep{Shakura1973}, $M_{\rm cnt}=M_1+M_2$ is the total mass of the central binary, and $H/r$ is the aspect ratio at the orbital distance of the planet.

\begin{figure}[t]
\centering
\includegraphics[width=0.8\linewidth]{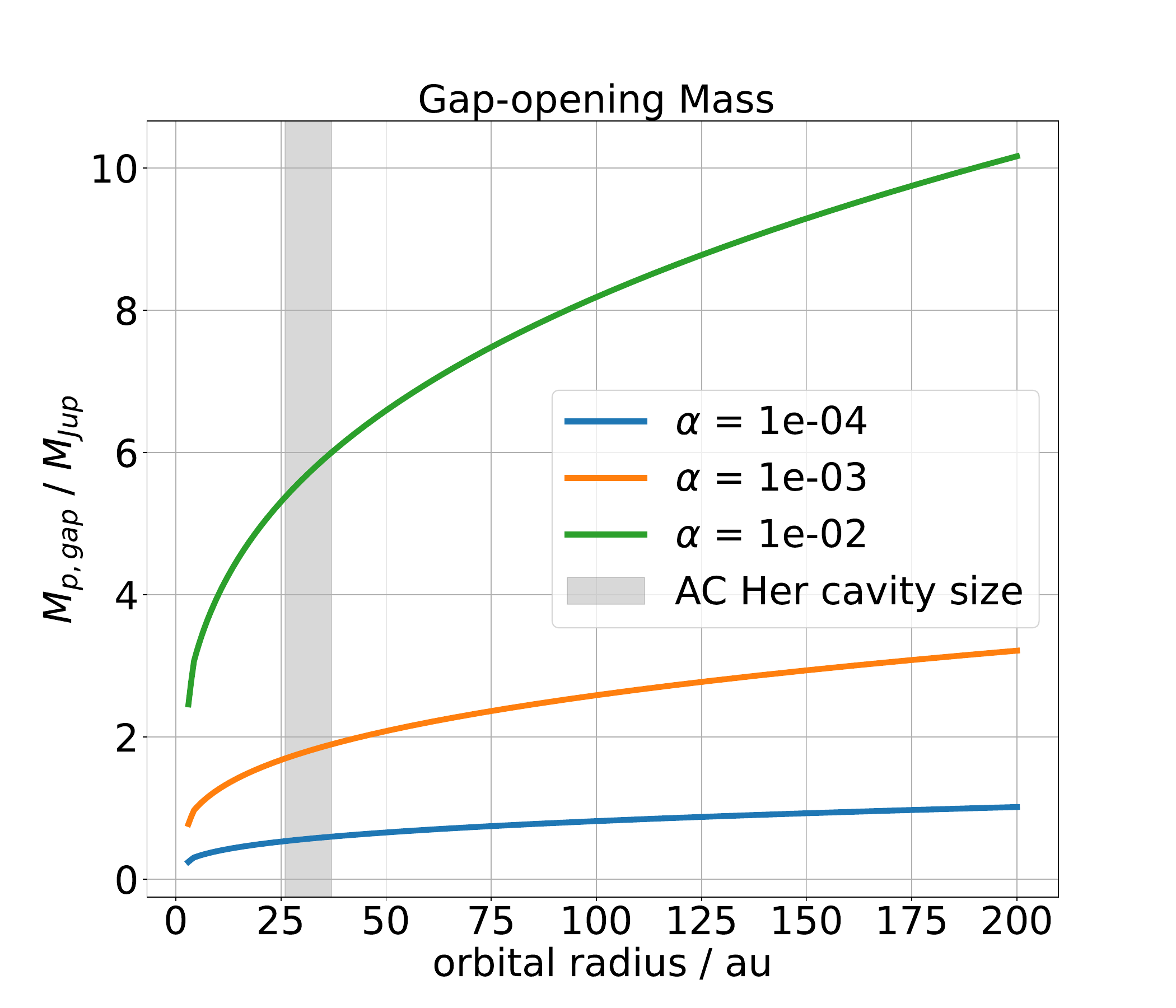}
\caption{Minimum mass for a planet to open a gap as a function of the planet's orbital parameter as calculated by Equation \eqref{eq:kanagawa}, for different values of $\alpha$. Shaded region indicates where observations have predicted the inner cavity of AC~ Her to extend to $R_{\rm in}=30_{-4}^{+7}$ au \citep{Anugu_2023} . This trend is independent of the disc mass.}
\label{fig:acher_cavity}
\end{figure}

Equation~\eqref{eq:kanagawa} shows that, for a fixed local disc structure, the gap-opening threshold depends on the local aspect ratio and viscosity, but not explicitly on the total disc mass or outer disc size. A second requirement is that the gap-opening timescale be shorter than the planet's migration timescale \citep{Malik_2015,1989ward}. For AC~Her, we find that this condition is always satisfied whenever the planet mass is larger than the gap opening mass. Figure \ref{fig:acher_cavity} has plotted the minimum gap-opening mass required at each distance in a post-AGB disc with the properties of AC~Her which are: $T_{\rm star}=6140$K \citep{mohorian2025tracingchemicaldepletionevolved}, $\log(L/L_\odot)=3.79$ (see Table \ref{tab:innercavities}), so $R_{\rm star}=69.4$R$_\odot$;  $M_{\rm disc} =2.5\times10^{-2}~$M$_\odot$ (assuming gas-to-dust ratio of 10 and the dust mass of $M_{\rm dust} =2.5\times10^{-3}~$M$_\odot$ taken from \citealt{Hillen_2015}),  $R_{\rm in}=30_{-4}^{+7}$ au \citep{Anugu_2023}, with an outer truncation radius of $R_{\rm out}=200$ au and the total binary mass is $M_{\rm cnt}=M_1+M_2=2.13~$M$_\odot$. We use Equation~\eqref{eq:aspectratio} to find the scale height $H$ and use the parametrisation of \citet{Shakura1973} for kinematic viscosity $\nu=\alpha c_s H$, where $\alpha$ is  a free parameter.

It is also possible to estimate the half-width of the gap created by a planet using the relation 
\begin{equation}
    \label{eq:gap_width}
    \frac{\Delta_{\rm Gap}}{R_p}= 0.41\left(\frac{M_p}{M_{\rm cnt}}\right)^{1/2}\left(\frac{H}{R_p}\right)^{-3/4}\alpha^{-1/4}, 
\end{equation}
\citep{Kanagawa_2016}. Using these two relations, we can argue that assuming $\alpha=10^{-3}$, a minimum planet mass of $8\,M_{\rm Jup}$ planet placed at $R_{\rm p}=15$~au in the disc of AC~Her produces a gap of half-width $\Delta_{\rm Gap}\approx10.5$~au, which is enough to extend outwardly to the observed edge of the inner cavity and inwardly to the expected dynamical truncation radius of $\sim9$~au (Table~\ref{tab:innercavities}). This demonstrates that the cavity in AC~Her could plausibly be explained by planet–disc interactions\footnote{Multiple planets can also carve gaps \citep[e.g.][]{Duffell_2015}, but require substantially more massive planets. We therefore neglect this scenario, as the limited solid mass budgets of post-AGB discs make such a population unlikely.}. We stress that this method places no constraints on how the planet formed.

\begin{table*}[t]
\resizebox{0.9\textwidth}{!}{%

\begin{threeparttable}
\caption{Inner cavity radii $R_{\rm cav}$ of the post-AGB discs found by \citet{Corporaal2023}, and values of  the sublimation radius, $R_{\rm subl}$ and the dynamical truncation radius, $R_{\rm dyn}$  of each of these systems. The properties we used to estimate the latter two radii (the luminosity of the primary star, $L_1$, the eccentricity, $e$, mass ratio, $q=M_1/M_2$, and orbital separation, $a$) are presented as well. For CT Ori, AD Aql, and ST Pup the quantities needed to estimate their dynamical truncation radii were not available in prior studies.}
\label{tab:innercavities}
\begin{tabular}{lccccccc}
\toprule

 \headrow System & $\log{\frac{L_1}{L_\odot}}$ 
& $e$
& $q$
& $a$
&$R_{\rm cav}$
& $R_{\rm subl}$
& $R_{\rm dyn}$ \\
 & - 
& - 
& -
& au
&au
& au
& au\\
\midrule

EP Lyr 
& $3.96^{a}$& $0.39^b $& $0.459^b$& $2.68^b$& $7$ 
& $5.07$
& $\approx3.7a=9.92$\\

RU Cen 
& $3.502^a$&$ 0.62 ^b$& $0.158^b$& $3.83 ^b$& $4$ 
& $3$
& $\approx3.2a=12.26$  \\

AC Her 
&  $3.39^c$, $4^d$, $3.79^a$& $0.206^c$ 
& $0.52^b$&$ 2.83^b$&  $25,30_{-4}^{+7}{}^{*}$ 
 & $2.64$,$5.31$,$4.18$ 
&  $9.03^e$   \\

AD Aql 
& $2.836^a$& -
& - 
& - 
& $18$ 
& $1.39 $
& -  \\

CT Ori 
& $3.260^a$& -
& - 
& - 
& $5$ 
& $2.27 $
& - 
\\

ST Pup 
& $2.964^a$&  $0.0^f$
& - 
& $a\sin{i}=0.67^f$
& $3$ 
& $1.61 $
& -  \\
\bottomrule
\end{tabular}
\begin{tablenotes}[hang]
\footnotesize
\item[] Notes: $^a$ \citet{Mohorian_2025b}, $^b$ \citet{Oomen_2020}, $^c$ \citet{2019Bodi},$^d$ \citet{2016Bertolami}, $^e$ \citet{Anugu_2023}, $^f$ \citet{Oomen_2018}. We note that \citet{Oomen_2018} also reported the temperature, mass function, and minimum mass of ST Pup, but the latter two properties are related to the inclination $i$, which is not determined in their study.
\item[] * The value $R_{\rm cav}=30_{-4}^{+7}$~au was reported by \citet{Anugu_2023}, which we use to estimate the gap-opening mass.
\end{tablenotes}
\end{threeparttable}%
}
\end{table*} 

\subsection{Grain Growth and Dust Evolution}
One of the primary marker of possible early planet formation is the presence of relatively large solid grains found in post-AGB systems. Dust coagulation to mm–cm sizes represents the first step toward planet formation. \citet{Scicluna_2020} measured spectral indices for a sample of post-AGB systems and found that grain growth to sizes of hundreds of micrometers is ubiquitous. They also reported not finding any correlation with disc structure and grain growth, which indicates that grain growth should have happened early enough in the these systems' lifetimes ($\ll 10^5$years). Individual systems from their sample are not listed in Table~\ref{tab:post-agb-list}, as the grain sizes are not constrained on a system-by-system basis.

Models of post-AGB discs around IRAS~08544-4431 \citep{Kluska_2018,Corporaal_2023}, AC~Her \citep{Hillen_2015} and the Red Rectangle \citep{bujarrabal2023compactdustdiskred,alcolea2023redrectanglediskbig}, use upper limits for the grain size distribution of approximately one millimetre, but are not necessarily constraining those values. However, the sub-millimetre fluxes observed by \citet{Hillen_2015} and \citet{Kluska_2018} do require millimetre-sized grains and cannot be explained by sub-micron particles alone.

\subsection{Radial Extent of the Dust and Gas Disc}

Comparing the radial extents of dust and gas discs provides a diagnostic of dust evolution. Because radial drift preferentially removes large grains from the outer disc, the ratio of the size of the gas disc to the dust disc can constrain the importance of grain growth and drift: ratios $\gtrsim4$ are indicative of significant grain growth and drift, while smaller ratios can also be caused by effects arising from the different behaviour between continuum and line optical depths \citep{2019facchini,Trapman_2019}.

In the Red Rectangle nebula, the post-AGB system hosts a compact dusty disc with a diameter of $\sim250$~au, embedded within a gas disc whose characteristic radius about 10 times larger \citep{bujarrabal2023compactdustdiskred}.

\subsection{Chemical Depletion}

Stellar surface chemical depletion characterised by the under-abundance of refractory elements (condensation temperatures $\gtrsim1250$~K) relative to volatile elements (condensation temperatures $\lesssim1250$~K) is observed in the atmospheres of nearly all post-AGB stars in these sytems and represents another similarity between post-AGB discs and some PPDs around YSOs \citep{1998Giridhar,mohorian2025tracingchemicaldepletionevolved} . 

The physical origin of depletion remains uncertain. A commonly invoked mechanism is the re-accretion of refractory-poor gas from the circumbinary disc, while refractory-rich dust is prevented from accreting by radiation pressure \citep{1992Waters}. However, simulations suggest this requires disc masses, lifetimes, and accretion rates higher than those typical of post-AGB systems \citep{Oomen_2019,martin2025modellingimpactcircumbinarydisk}. In PPDs around YSOs, analogous gas-dust separation has been attributed to photoevaporation, dead zones, grain growth, or giant planets \citep{mohorian2025tracingchemicaldepletionevolved}.

\section{Growth Barriers in Dust Coagulation}
\label{ap:barriers}
Here, we discuss the three main barriers to grain growth via collisions, and describe the equations used to delineate the caps on grain size in post-AGB discs in Figure \ref{fig:grains_pagb}. 

All three barriers computed rely on a method proposed by \citet{Birnstiel_2012} to characterise the largest grain size $a$ attainable through collisions for the different barriers by relating the Stokes number $S_{\rm t}$ to the collision threshold velocity $v_{\rm thr}$

\begin{equation}
S_{\rm t} = \frac{v_{\rm thr}^2}{3\alpha c_{\rm s}^2}.
\label{eq:stokes_threshold}
\end{equation}

Here, $\alpha$ is the turbulence parameter and $c_{\rm s}$ is the sound speed. The threshold velocity $v_{\rm thr}$ depends on the growth barrier under consideration, as detailed in the following subsections.

\subsection{The Bouncing Barrier}
\label{ssec:the_bouncing_barrier}
The bouncing barrier arises because compact particles beyond a certain size tend to bounce off each other during collisions rather than sticking and growing, as demonstrated in laboratory experiments such as those by \citet{2000blum_experiments}.

\citet{dominik2023bouncingbarrierrevisitedimpact} use the framework of \citet{Birnstiel_2012}\footnote{Other studies have characterised the bouncing barrier differently, e.g., \citet{drazkowska2023planetformationtheoryera} use a terminal velocity of $1$~cm.s$^{-1}$, while \citet{2017Johansen} impose a limit on the Stokes number.} to find the maximum grain size achievable due to the bouncing barrier $a_b$ to be
\begin{equation}
\label{eq:a_bounce}
    a_b=max(a_\eta,a_{b_0}),
\end{equation}
where
\begin{equation}
\label{eq:a_b0}
   a_{b_0}=\left(\frac{5}{\pi}\frac{\Sigma_ga_{\rm mono}F_{\rm roll}}{\alpha c_s^2 \rho_s^2}\right)^{0.25}, 
\end{equation}
where $F_{\rm roll}=10^{-4} \rm dyn$ is the force needed to roll a monomer in contact over another monomer, $a_{\rm mono}$ is the size of the monomers that make up the aggregates (usually set to $a_{\rm mono}=1\mu m$), and $a_\eta$ is a term to account for the truncation of the Kolmogorov turbulent cascade at small scales \citet{dominik2023bouncingbarrierrevisitedimpact}
\begin{equation}
\label{eq:a_kolmogorov}
    a_\eta=\frac{\rho_g c_s}{\rho_s\Omega_K}\sqrt{\frac{8}{\pi Re}}.
\end{equation}
Here, $Re$ is the Reynolds number which is
\begin{equation}
    Re=\frac{\alpha c_s H}{\nu_{\rm mol}},
\end{equation}
where $\nu_{\rm mol}=\lambda_{\rm mfp}u_{\rm th}/2$ is the molecular viscosity for molecular hydrogen, where $\lambda_{\rm mfp}$ is the mean free path and $u_{\rm th}$ is the thermal velocity (for further details refer to \citealt{dominik2023bouncingbarrierrevisitedimpact}). \citet{dominik2023bouncingbarrierrevisitedimpact} find that the bouncing barrier narrows the grain size distribution, eliminating most $\mu$m-sized grains. They argue that this limits the maximum attainable Stokes number, slowing the streaming instability growth, though continuous size distributions can also hinder the streaming instability \citep{2019Krapp}.

\subsection{The Fragmentation Barrier}
The fragmentation barrier is typically characterised by a threshold velocity above which particles break apart and regrow cyclically. While this threshold depends on particle properties \citep{2000blum_experiments}, most simulations adopt a constant value for simplicity \citep{drazkowska2023planetformationtheoryera}. Reported values range from $\sim 1\, \rm m\, s^{-1}$ for silicates \citep{2010guttler_zoo} while \citet{2014yamamoto} reported values from $30\,\mathrm{m\,s^{-1}}$ for silicates to $80\, \mathrm{m\,s^{-1}}$ for porous icy aggregates. \citet{Birnstiel_2012} report the maximum particle size set by the fragmentation barrier to be
\begin{equation}
    \label{eq:fragment_maxsize}
a_{\rm frag}= f_f\frac{2\Sigma_{\rm g}v_{\rm f}^2}{3\pi \rho_{\rm s}\alpha c_{\rm s}^2},
\end{equation}
where $f_f$ is an offset factor (numerically found to be $f_f=0.37$).

\subsection{The Drift Barrier}
\label{sec:drift}
The drift barrier is caused by a headwind being exerted on approximately Keplerian solid particles by the slightly sub-Keplerian gas in a disc, which becomes more significant for larger particles as their drift timescales approach their growth timescales. \citet{Birnstiel_2012} propose the following expression for the maximum particle size (in the Epstein regime)
\begin{equation}
    \label{eq:drift_maxsize}
a_{\rm drift}= f_d\frac{2\Sigma_{\rm d}v_{\rm K}^2}{\pi \gamma \rho_{\rm s}c_{\rm s}^2},
\end{equation}
where $f_d=0.55$ is a correction factor, $v_{\rm K}$ is the Keplerian velocity, and $\gamma$ is the power-law index of the gas pressure profile ($P=\rho^\gamma$).

\end{document}